\documentclass[prd,aps,reprint,10pt,nofootinbib,superscriptaddress]{revtex4-1}
\usepackage{amsmath}
\usepackage{cases}
\usepackage{amssymb}
\usepackage{scalerel}
\usepackage{stackengine,wasysym}
\usepackage{mathrsfs}
\usepackage{graphicx}
\usepackage{color}

\newcommand\reallywidetilde[1]{\ThisStyle{%
		\setbox0=\hbox{$\SavedStyle#1$}%
		\stackengine{-.1\LMpt}{$\SavedStyle#1$}{%
			\stretchto{\scaleto{\SavedStyle\mkern.2mu\AC}{.5150\wd0}}{.6\ht0}%
		}{O}{c}{F}{T}{S}%
}}

\begin{document}

	\title{Quantum Kasner transition in a locally rotationally symmetric Bianchi II universe}
	
	\author{Ana Alonso-Serrano} 
	\email{ana.alonso.serrano@aei.mpg.de}
	\affiliation{Max-Planck-Institut f\"ur Gravitationsphysik 	(Albert-Einstein-Institut), Am M\"uhlenberg 1, 14476 Potsdam, Germany}
	\author{David Brizuela}
	\email{david.brizuela@ehu.eus}
	\affiliation{Fisika Saila, Universidad del Pa\'is Vasco/Euskal Herriko Unibertsitatea (UPV/EHU), Barrio Sarriena s/n, 48940 Leioa, Spain}
	\author{Sara F. Uria}
	\email{sara.fernandezu@ehu.eus}
	\affiliation{Fisika Saila, Universidad del Pa\'is Vasco/Euskal Herriko Unibertsitatea (UPV/EHU), Barrio Sarriena s/n, 48940 Leioa, Spain}
	
\date{\today}

\begin{abstract}

The Belinski-Khalatnikov-Lifshitz (BKL) conjecture predicts a chaotic alternation of Kasner epochs in the evolution of
generic classical spacetimes towards a spacelike singularity. As a first step towards understanding the full quantum BKL
scenario, we analyze a vacuum Bianchi II model with local rotational symmetry, which presents just one Kasner transition.
During the Kasner epochs, the quantum state is coherent and it is thus
characterized by constant values of the different quantum fluctuations, correlations
and higher-order moments. By computing the constants of motion of the system we provide,
for any peaked semiclassical state, the explicit analytical transition rules that relate
the parametrization of the asymptotic coherent state before and after the transition.
In particular, we obtain the modification of the transition rules for the classical variables
due to quantum back-reaction effects.
This analysis is performed by considering a high-order truncation in moments (the full computations are performed up to
fifth-order, which corresponds to neglecting terms of an order $\hbar^3$),
providing a solid estimate about the quantum modifications to the classical model.
Finally, in order to understand the dynamics of the state during the transition, we perform
some numerical simulations for an initial Gaussian state, that show that the initial and
final equilibrium values of the quantum variables are connected by strong and rapid oscillations.
\end{abstract}

\keywords{}

\maketitle

\section{Introduction}

It has been conjectured that, close to a spacelike singularity, the classical dynamics of any universe
follows a chaotic behavior given by the Belinski-Khalatnikov-Lifshitz (BKL) scenario~\cite{BKL}.
According to this,
different points decouple and the dynamics of each point can be described by
a Bianchi IX universe, which is characterized by an alternation of different Kasner epochs
(corresponding to a Bianchi I vacuum or Kasner solution) connected by quick transitions
(as compared with the duration of the epochs). The alternation among epochs implies that different (anisotropic) spatial
directions are squeezed and stretched during the evolution towards the singularity, resulting in the mentioned chaotic behavior.
One relevant characteristic of this conjecture is that this behavior is dominated by the vacuum solution, and the introduction
of matter (besides particular cases) does not change the qualitative behavior. 

The analysis of Bianchi IX models can be very complicated and it is often convenient
to describe the system in terms of the Misner variables~\cite{Misner,Misner2},
in the context of the so-called \textit{Mixmaster Universe};
that is, instead of using the scale factors, the system is characterized in terms of the spatial volume and two anisotropic shape-parameters.
The dynamics of these models is then described as a free particle with a potential that drives the transitions between Kasner regimes
via exponential walls~\cite{Misner,Maccallum,Bojowald:2006da}, in the so-called ``cosmological billiard''~\cite{Damour:2002et,Heinzle:2007kv}.
A simpler scenario corresponds to a Bianchi II universe. This model presents only one transition between two different Kasner epochs
on its evolution towards the singularity. Nonetheless, any Bianchi IX spacetime can be understood as a succession of
Bianchi II models providing transitions among different Kasner epochs~\cite{Heinzle:2009eh}.

This picture of the classical dynamics remains being a conjecture, although it is widely accepted and numerous numerical studies confirm the described behavior \cite{Berger:2002st,Garfinkle:2003bb,Heinzle:2012um}.
But let us remark that the BKL scenario takes place very close to the singularity.
Therefore, at this stage, one would
expect quantum effects to become relevant in the analysis. However, it is still unclear how this scenario can be modified when
quantum effects are taken into account.  In the literature, there are several approaches to quantum models of Bianchi IX, focused on different features and questions (as the avoidance of the singularity or the survival of the chaotic behavior)~\cite{Berger:1989jm,Bojowald:2004ra,Kheyfets:2005nz,Benini:2006xu,Wilson2010,Ashtekar:2011ck,Bergeron:2015ppa,Czuchry:2016rlo,Wilson-Ewing:2018lyx,Kiefer:2018uyv,Gozdz:2018aai,Giovannetti:2019ewe}, as well as simpler approaches to Bianchi II quantum models~\cite{Ashtekar:2009um,Corichi:2012hy,Bergeron:2014kea,Saini:2017ipg}. 

In particular, one question concerns the survival of the structure
of Kasner epochs connected by quick transitions,
and whether they follow the same transition rules as in the classical picture.
In this paper, we analyze this question within the context of a simple \textit{locally rotationally symmetric} (LRS) Bianchi II model.
We first analyze the canonical structure and exact solutions of the classical model,
identifying the transition rules of the canonical variables of the system.
Then we develop an exact quantization of the model and
focus on analyzing the survival of
the Kasner transition and the quantum modification of the transition rules.
For such a purpose, we will perform a decomposition of the wavefunction into its
infinite set of moments following the framework first presented in \cite{Bojowald:2005cw}.
This formalism is very useful to understand the dynamics of peaked semiclassical states
and it has already been applied to the analysis of a variety of cosmological models; see, e.g., \cite{Bojowald:2020emy}--\cite{Bojowald:2010qm}.

The rest of the paper is organized as follows. In Sec.~\ref{sec:class}
the canonical structure of the classical model is described in terms of Misner variables
and the classical transition rules are obtained.
The full quantum analysis is then performed in Sec.~\ref{sec:quant}, which is divided into
four subsections. In Subsec. \ref{subsec:set_up} the quantum
moments and their equations of motion are derived, whereas in Subsec. \ref{susbec:conserved_quant}
the constants of motion of the system are constructed.
Making use of these constants of motion, in Subsec. \ref{sec_quantumtransition} the quantum transition rules for any semiclassical
peaked state are obtained. Subsec.~\ref{sec_numerics} completes the study
with a numerical simulation that describes in detail the dynamics during the transition.
The discussion of the results and possible future developments are finally presented in Sec.~\ref{sec:discuss}.

\section{Canonical analysis of the classical system} \label{sec:class}

The metric of a general homogeneous but anisotropic universe can be expressed as follows
\begin{equation}\label{line_element}
	ds^2=-N^2dt^2+\sum_{i,j,k=1}^3a_k^2l^k_il^k_j dx_idx_j,
\end{equation}
where $N$ is the lapse function, $a_k$ is the scale factor in the $\vec{l}^{k}=(l_1^k,l_2^k,l_3^k)$
spatial direction, and the $l^k_i$ vectors (or Kasner axes) form an orthonormal triad that determines
the spatial direction of the anisotropic expansion or contraction of the universe.
For the Bianchi II model, these vectors
verify the following conditions,

\begin{eqnarray*}
\sum_{i,j=1}^3(l_{i,j}^k-l_{j,i}^k)l^i_nl^j_m=\delta_{k,3}\epsilon_{knm},
\end{eqnarray*}
with $\epsilon_{knm}$ being the complete antisymmetric Levi-Civita symbol.

A very convenient way to describe this spacetime is given by the variables introduced by Misner \cite{Misner,Misner2}.
The three different scale factors are written as
\begin{equation}\label{scale_factors_misner_var}
	a_k=e^{\alpha+\beta_k},
\end{equation}
where $\alpha$ encodes the spatial volume, $e^{\alpha}=(a_1a_2a_3)^{1/3}$, and the variables $\beta_k$ satisfy the constraint
$\beta_1+\beta_2+\beta_3=0$. Therefore, the following two independent shape-parameters can be defined
\begin{eqnarray}\label{b_plus_b_minus}
	\beta_+&:=&-\frac{1}{2}\beta_3=-\frac{1}{2}\ln
	\Bigg[\frac{a_3}{(a_1a_2a_3)^{1/3}}\Bigg], \nonumber \\[7pt]
	\beta_-&:=&\frac{1}{2\sqrt{3}}(\beta_1-\beta_2)=
\frac{1}{2\sqrt{3}}\ln
	\bigg(\frac{a_1}{a_2}\bigg).
\end{eqnarray}
As one can directly check, the range of $\alpha$ and $\beta_{\pm}$ corresponds to the whole real line.

In this paper we will focus on the analysis of the vacuum case. For this Bianchi II vacuum model, the
Hamiltonian constraint reads
\begin{equation}\label{constraint}
	{\cal C}=\frac{1}{2} e^{-3 \alpha}
   \left(-p_\alpha^2+p_{-}^2+p_{+}^2\right)+e^{\alpha} U(\beta_+,\beta_-)=0,
\end{equation}
where $p_{\alpha}=-\frac{e^{3\alpha}}{N} \frac{d\alpha}{dt}$ and $p_{\pm}=\frac{e^{3\alpha}}{N} \frac{d\beta_{\pm}}{dt}$ are the conjugate momenta of $\alpha$ and $\beta_{\pm}$, respectively, and the potential term takes the value $U(\beta_+,\beta_-)=e^{-8\beta_+}$.
Other Bianchi models are described by this very same Hamiltonian, but with a different form of the potential.
In particular, the vacuum Bianchi I --or Kasner-- model, which will be relevant in the subsequent discussion,
is given by $U(\beta_+,\beta_-)=0$.

Since it is a monotonic function,
in these models it is usual to choose $\alpha$ as the internal time. The above constraint is then
deparametrized to define the physical Hamiltonian
\begin{eqnarray}\label{hamiltonian}
 H:=&&-p_\alpha=\big[p_+^2+p_-^2+2e^{4\alpha}U(\beta_+,\beta_-)\big]^{1/2}
 \nonumber\\[5pt]= &&\big(p_+^2+p_-^2+2e^{4\alpha
 	-8\beta_+}\big)^{1/2}.
\end{eqnarray}

In order to understand and study this system analytically in detail, we will introduce a further simplifying assumption,
and focus on models with a preferred spatial direction, which are known as \textit{locally rotationally symmetric} (LRS) spacetimes\footnote{These spacetimes have been largely studied in cosmology; check, \textit{e.g.}, ~\cite{Ellis1967,vanElst:1995eg,Clarkson:2002jz,Betschart:2004uu,Clarkson:2007yp,Singh:2016qmr,Khan:2016dqj,Bergh:2017sps,Singh:2017qxi,Bojowald:2003md}.}. In this case, by choosing the third direction as the preferred one,
this corresponds to assuming $a_1=a_2$ and thus the shape-parameter $\beta_-$, as well as its conjugate momentum $p_-$,
are vanishing: $\beta_-=0=p_-$. Therefore, defining for compactness $\beta:=\beta_+$ and $p:=p_+$,
in this LRS model
the Hamiltonian (\ref{hamiltonian}) is reduced to
\begin{equation}\label{hamiltonian_LRS}
H=\big(p^2+2e^{4\alpha}U\big)^{1/2}=\big(p^2+2e^{4\alpha-8\beta}\big)^{1/2}.
\end{equation}
This is a system with just one degree of freedom $(\beta,p)$, and
the equations of motion for the canonical pair can readily be obtained,
\begin{eqnarray}
\label{eq_m_beta}
 \dot{\beta}&=&
 \frac{p}{H},
  \\[2pt]
 \label{eq_m_p}
\dot{p}&=& \frac{8}{H}e^{4\alpha}U=
 \frac{8}{H}e^{4\alpha-8\beta},
\end{eqnarray}
where the dot represents the derivative with respect to the internal time $\alpha$.

The regions where the potential term $e^{4\alpha}U$ has a negligible contribution to the Hamiltonian $H$,
that is, where the dimensionless ratio $r=2e^{4\alpha}U/p^2$ is very small $r\ll 1$,
are the so-called Kasner regimes or epochs. (Note that, as commented above, in the Kasner model the potential
is exactly vanishing $U=0$ and thus also the corresponding dimensionless ratio $r$.)
In these regimes, the equations of motion can easily be solved. On the one hand,
$p$ is a constant of motion and, in this sense, the system follows a free dynamics.
On the other hand, the shape-parameter $\beta$ is a linear function of $\alpha$
\begin{eqnarray}
\label{beta_Kasner}
	\beta&=&\text{sign}(p)\alpha+c,
\end{eqnarray}
with the integration constant $c$.
Note that $\beta$ either grows or decreases according to the sign of the variable $p$.
Therefore, each Kasner epoch is completely characterized by the values of the constants of motion
$p$ and $c$.

In fact, the dynamics of the Bianchi II model can be understood as two (asymptotic) Kasner epochs
and a transition between them. Let us describe this in more detail.
We are interested in a model corresponding to an expanding universe with a singularity in the past $\alpha\to -\infty$.
Our analysis will begin on a macroscopic universe far away from the singularity,
that is, with a large initial value of $\alpha$. If the initial values of $\beta$ and $p$ are chosen in such
a way that the ratio $r$ is small, the system will be on a Kasner epoch.
Depending on the sign of $p$, $\beta$ will be an increasing or decreasing function \eqref{beta_Kasner}. For definiteness, and without
loss of generality, let us choose the initial $p$ as positive.\footnote{If the initial $p$ is
chosen as negative, the system will follow the same qualitative behavior as described, but with the direction of time inverted.}
In this case, towards larger values of $\alpha$, the shape parameter $\beta$ will be growing,
making the ratio $r$ more and more negligible. Therefore, in this direction, the Kasner condition holds
and the dynamics is given by \eqref{beta_Kasner} up to $\alpha\rightarrow\infty$.
On the contrary, towards lower values of $\alpha$, the shape-parameter $\beta$ will be decreasing and
thus the ratio $r$ increasing. Therefore, it will reach a point where $r$ will be no longer negligible
and the Kasner approximation will break down. From this point on, $p$ is no longer constant and
the dynamics of the shape-parameter can not be approximated by \eqref{beta_Kasner} anymore.

\begin{figure}[h]
	\centering
	\includegraphics[width=0.85\linewidth]{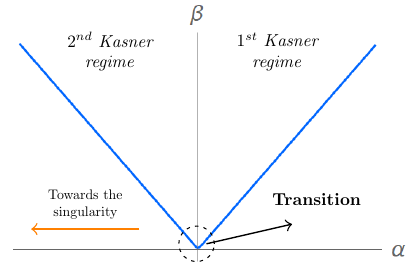}
	\caption{Classical evolution of the variable $\beta$ with respect to $\alpha$, where we have chosen $p>0$ for large values of $\alpha$ and $\beta$.}
	\label{fig:beta}
\end{figure}

\begin{figure}[h]
	\centering
	\includegraphics[width=0.85\linewidth]{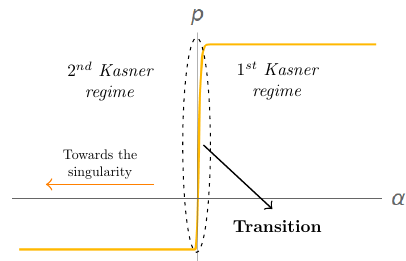}
	\caption{Classical evolution of the variable $p$ with respect to $\alpha$, where we have chosen $p>0$ for large values of $\alpha$ and $\beta$.}
	\label{fig:p}
\end{figure}

Nonetheless, this period is very short. A very quick transition happens, which changes the sign
of $p$ and produces a bounce in $\beta$, and then the system enters into another Kasner regime
with corresponding parameters $\widetilde c$ and $\widetilde p<0$.
This new Kasner epoch evolves towards lower values of $\alpha$ until reaching the singularity at $\alpha \rightarrow -\infty$.
During this evolution, the shape parameter $\beta$ increases linearly, following \eqref{beta_Kasner}, and tending to infinity at the singularity.
For the sake of clarity, the evolution of $\beta$ and $p$ is illustrated in Figs. \ref{fig:beta} and \ref{fig:p} respectively.


From now on, we will focus on the study of this transition, specifically on the rules that relate the parameters
that characterize the first Kasner epoch $(\overline c, \overline p)$, which will be denoted with a bar,
with those that describe the second one $(\widetilde c, \widetilde p)$, and will be denoted with a tilde. The seminal idea to perform this analysis~\cite{Misner} is to find the constants of motion of the system. Since they are conserved all along the evolution, these quantities shall be the same for both Kasner epochs. Therefore, one can write them in terms of the corresponding parameters
in each Kasner epoch, before ($\alpha\to+\infty$) and after ($\alpha\to-\infty$) the transition, and obtain the relations
$\widetilde c=\widetilde c(\overline c,\overline p)$ and $\widetilde p=\widetilde p(\overline c,\overline p)$. These relations will be known as the transition rules.

The first nontrivial task is thus to find the constants of motion. Any conserved quantity $R=R(\alpha,\beta,p)$
must obey the following equation
\begin{eqnarray}
0=\frac{d R}{d\alpha}&=&
\frac{\partial R}{\partial\alpha}+\frac{\partial R}{\partial\beta}\frac{d\beta}{d\alpha}+\frac{\partial R}{\partial p}\frac{dp}{d\alpha}
\nonumber\\[5pt]
&=&
\frac{\partial R}{\partial\alpha}+\frac{\partial R}{\partial\beta}\frac{\partial H}{\partial p}-\frac{\partial R}{\partial p}\frac{\partial H}{\partial\beta},
\end{eqnarray}
where we have made use of definitions (\ref{eq_m_beta})-(\ref{eq_m_p}). In order to find the solution,
it is convenient to perform the following transformation $\alpha\rightarrow H$, so that $R=R(H,\beta,p)$,
and the above equation reads
\begin{equation}
0= \frac{\partial R}{\partial H}\frac{\partial H}{\partial\alpha}+\frac{\partial R}{\partial\beta}\frac{\partial H}{\partial p}-\frac{\partial R}{\partial p}\frac{\partial H}{\partial\beta}.
\end{equation}
Replacing in this expression the form of the Hamiltonian (\ref{hamiltonian_LRS}) one obtains,
\begin{equation}
0=2(H^2-p^2)\frac{\partial R}{\partial H}+\frac{\partial R}{\partial\beta} p
+4(H^2-p^2)\frac{\partial R}{\partial p}.
\end{equation}
The general solution to this equation yields
\begin{align}
R&=
F\Big[2H-p,
e^{2(\alpha+\beta)}(H-p)
\Big],
\label{conserved_quant_general}
\end{align}
for any generic function $F$. Consequently, we will consider the following two independent conserved quantities
to analyze the transition
\begin{eqnarray}\label{conserved_quantities}
	R_1&:=&2H-p,\\
	R_2&:=&e^{2(\alpha+\beta)}(H-p).
\end{eqnarray}

In the Kasner regimes, where the ratio is very small $r\ll1$, the conserved quantities are approximately given by
\begin{align}\label{conserved_quantities_Kasner}
R_1&\approx
2|p|-p,
\\[10pt]
R_2&\approx e^{2(\alpha+\beta)}\Bigg[
|p|\bigg(
1+\frac{e^{4\alpha-8\beta}}{p^2}
\bigg)-p
\Bigg].
\end{align}
By considering the functional form for $\beta$ \eqref{beta_Kasner}
in both asymptotic regions, and taking into account that the initial value of $p$ is chosen as positive,
$\overline p>0$, the requirement of conservation of the above quantities provides the following
two equations,
\begin{eqnarray}
&& \overline p=[2\text{sign}(\widetilde{p})-1]\widetilde{p},
\label{Kasner_equateR1}
\\
&&\dfrac{e^{-6 \overline c}}{\overline p}=
\widetilde{p}e^{2[1+\text{sign}(\widetilde{p})]\alpha+2\widetilde{c}} 
\label{Kasner_equateR2}
\\
&&\hspace{1.2cm}\times
\Bigg[
\text{sign}(\widetilde{p})\bigg(
1+\dfrac{e^{4[1-2\text{sign}(\widetilde{p})]\alpha-8\widetilde{c}}}{\widetilde{p}^2}
\bigg)-1
\Bigg],
\nonumber
\end{eqnarray}
which relate the parameters of the first Kasner epoch $(\overline c, \overline p)$
with those of the second Kasner regime $(\widetilde c, \widetilde p)$.
It is then straightforward to solve this system and obtain the classical transition rules
\begin{align}
\label{transition_p}
	&\widetilde{p}=-\frac{1}{3}\overline p,
	\\
	&\widetilde{c}=-3 \overline{c}-\frac{1}{2}\ln
	\bigg(
	\frac{2}{3}\overline{p}^3
	\bigg).
	\label{transition_beta}
\end{align}

\section{Quantum analysis} \label{sec:quant}

Once we have provided a canonical description of the classical system and the classical results are under control,
we are ready to develop the quantum analysis. The present section is divided into four subsections. In Subsec.
\ref{subsec:set_up} we will define the quantum moments, which will be the basic variables to describe the quantum states.
Their equations of motion are obtained and solved in the asymptotic Kasner regimes. In particular, it is shown that,
for semiclassical states, the qualitative dynamical picture of the system is kept as in the classical case: namely,
there will be two different Kasner epochs with a transition connecting them.
Following the classical analysis, in Subsec. \ref{susbec:conserved_quant}, the conserved quantities for the quantum system will be derived.
These constants of motion will be then used to obtain analytically the quantum transition rules between the two Kasner
epochs in Subsec. \ref{sec_quantumtransition}.
This will provide the asymptotic behavior of all the quantum moments for any initial state. Nonetheless, in order to
understand better the dynamics during the transition, in Subsec. \ref{sec_numerics} a numerical implementation of the system of equations is performed
for the particular case of a Gaussian initial state.

\subsection{Set up}
\label{subsec:set_up}

The quantum dynamics of this model is governed by a Hamiltonian $\hat{H}$, which is obtained
by promoting the classical Hamiltonian (\ref{hamiltonian_LRS}) to an operator.
We will analyze the dynamics by using a formalism based on a moment decomposition of the wave
function~\cite{Bojowald:2005cw}. More precisely, we define the quantum moments as
\begin{equation}
\label{moments}
	\Delta(\beta^ip^j):=\Big\langle
	(\hat{\beta}-\langle
	\hat{\beta}\rangle
	)^i(\hat{p}-\langle
	\hat{p}\rangle
	)^j
	\Big\rangle_{\text{Weyl}},
\end{equation}
where $i,j\in\mathbb{N}$, $\langle\,\rangle$ indicates the expectation value and
the Weyl subscript refers to a totally symmetric ordering of the operators. This infinite set of moments, together with $\langle\hat{\beta}\rangle$ and $\langle\hat{p}\rangle$, fully characterizes the quantum state.
The sum of the indices $i+j$ will be referred as the \textit{order} of the corresponding moment. This order will
be later used to introduce a cut-off on this infinite system.

The dynamics of the moments (\ref{moments}) is driven by a quantum effective Hamiltonian, that is defined as the expectation value
of the operator $\hat{H}$. By performing a Taylor expansion around the expectation values of $\hat{\beta}$ and $\hat{p}$,
this effective Hamiltonian can be written as follows,
\begin{align}
 H_Q&=
 \big\langle
 \hat{H}(\hat{p},\hat{\beta})
 \big\rangle
 \label{effective_hamiltonian}
 \\
 &=H+
\sum_{i+j=2}^{\infty} \frac{1}{i!j!}
 \dfrac{\partial^{i+j}H(\beta,p)}{\partial\beta^i\partial p^j}\Delta(\beta^ip^j)
 ,
  \nonumber
\end{align}
where $\beta:=\langle\hat{\beta}\rangle$ and $p:=\langle\hat{p}\rangle$ have been defined,
and $H(\beta,p)$ is the classical Hamiltonian (\ref{hamiltonian_LRS}). As it can be seen, the Hamiltonian operator has been chosen as Weyl-ordered.

The equations of motion for the variables of the system ($\beta$, $p$ and $\Delta(\beta^ip^j)$) can then be
obtained by computing their Poisson brackets with the effective Hamiltonian $H_Q$:
\begin{align}
\label{quantum_em_beta}
	&\frac{d\beta}{d\alpha}
	=\{
	\beta,H_Q
	\}=\frac{\partial H_Q}{\partial p},
	\\
\label{quantum_em_p}
&\frac{dp}{d\alpha}
=\{
p,H_Q
\}=-\frac{\partial H_Q}{\partial\beta},
\\
&\frac{d\Delta(\beta^ip^j)}{d\alpha}
=\{
\Delta(\beta^ip^j),H_Q
\}
\label{quantum_em_G_ij}
\\
\nonumber
&\hspace{30pt}=
\sum_{m,n=0}^{+\infty} \frac{1}{m!n!}
\dfrac{\partial^{m+n}H(\beta,p)}{\partial\beta^m\partial p^n}\{\Delta(\beta^ip^j),\Delta(\beta^mp^n)\},
\end{align}
where $\{\,,\,\}$ are the Poisson brackets between expectation values defined as $\{\langle{\hat{f}}\rangle,\langle{\hat{g}}\rangle\}=\frac{1}{i\hbar}\langle[{\hat{f},\hat{g}}]\rangle$, $\hat{f}$ and $\hat{g}$ being arbitrary operators. Moreover, the general expression for the brackets between any two
moments is known \cite{Bojowald:2005cw,Bojowald:2010qm} to be
\begin{widetext}
\begin{align*}
	\{\Delta(\beta^kp^l),
	\Delta(\beta^mp^n)\}=
	ml\Delta(\beta^kp^{l-1})
	\Delta(\beta^{m-1}p^n)
	-
	kn\Delta(\beta^{k-1}p^{l})
	\Delta(\beta^{m}p^{n-1})
	+\sum_{\substack{r=1\\[2pt]\text{odd}}}^{S}
	\bigg(\frac{i\hbar}{2}\bigg)^{r-1}K_{klmn}^{r}
	\Delta(\beta^{k+m-r}p^{l+n-r}),
\end{align*}
\end{widetext}
where $S:=\min(k+l,m+n,k+m,l+n)$ and the coefficients $K^r_{klmn}$ are defined as
\begin{align*}
	K^r_{klmn}=\sum_{s=0}^{r}
	(-1)^ss!(r-s)!
	\binom{k}{r-s}
	\binom{l}{s}
		\binom{m}{s}
	\binom{n}{r-s}.
\end{align*}

Note that, in general, \eqref{quantum_em_beta}--\eqref{quantum_em_G_ij} form an infinite
system of coupled equations. Therefore, in order to analyze the system, one usually
introduces a truncation by considering negligible all the moments $\Delta(\beta^np^m)$
with an order $n+m$ higher than a certain cut-off $N$. A moment $\Delta(\beta^np^m)$
has the dimensions of $\hbar^{(n+m)/2}$ and, in this sense, this kind of truncation
is valid during semiclassical regimes, that is, as long as the state remains peaked
on a classical trajectory.

Nonetheless, if the Hamiltonian is at
most quadratic on the basic variables, the different orders decouple and one usually is
able to solve the complete infinite set of equations. In fact, this is the case in
the Kasner epochs analyzed above (when the ratio $r=2e^{4\alpha-8\beta}/p^2$ is negligible),
as then the classical Hamiltonian is linear in $p$. In this approximation, the equations of motion take the form,
\begin{eqnarray*}
	&&\frac{d\beta}{d\alpha}\approx \text{sign}(p), \hspace{0.3cm} \frac{dp}{d\alpha}\approx 0,\hspace{0.3cm} \frac{d\Delta(\beta^{i}p^{j})}{d\alpha} \approx 0,
\end{eqnarray*}
which are immediate to solve
\begin{eqnarray}\label{quantum_Kasner_beta}
&&\beta\approx \text{sign}(p)\alpha+c, 
\\[5pt]
\label{quantum_Kasner_p}
&& p\approx const.,
\\[5pt]
\label{quantum_Kasner_G_ij}
&& \Delta(\beta^{i}p^{j})\approx const.,
\end{eqnarray}
with the integration constant $c$. As can be seen, the dynamics during the Kasner regime
is that of a coherent state. There is no quantum back-reaction, as the moments are constants of motion
and evolve decoupled from the expectation values, and the behavior of the expectation values remains as in the classical case. In summary,
the parameters that characterize each quantum Kasner epoch will be, as in the classical case, the constants $c$ and $p$,
along with the infinite constant set of  moments $\Delta(\beta^{i}p^{j})$.

Therefore, our goal is to obtain the quantum transition rules, that relate the set of parameters that
describe the initial Kasner epoch $(\overline c, \overline p, \overline{\Delta(\beta^{i}p^{j})})$, which will
be denoted with a bar, with the parametrization of the second Kasner regime $(\widetilde c, \widetilde p,\reallywidetilde{\Delta(\beta^{m}
p^{n})})$, denoted with a tilde. In this way, we will be able to see how the classical transition rules,
described by equations (\ref{transition_p})-(\ref{transition_beta}), are modified by quantum effects.
For such a purpose, let us obtain the constants of motion for the quantum system.

\subsection{Conserved quantities}
\label{susbec:conserved_quant}

In order to obtain the constants of motion, we first note that, since we are dealing with an
infinite system of equations, we will need an infinite amount of them. Nonetheless, if
two independent conserved operators $\hat{R}_1$ and $\hat{R}_2$ are known, their expectation
values $\langle\hat{R}_1\rangle$ and $\langle\hat{R}_2\rangle$ would be trivially conserved.
In fact, one would be able to generate an infinite set of conserved quantities, just by taking the expectation
values of any product between them, that is,
\begin{equation}
\label{quantum_const_motion}
 \langle \hat{R}_i^n\hat{R}_j^m \rangle=const., \hspace{0.2cm}
 \text{for}
 \hspace{0.2cm}m,n\in\mathbb{N}\;\text{and}\; i,j\in\{1,2\}.
\end{equation}

Obtaining the operators $\hat R_1$ and $\hat R_2$ might be a very complicate task, so we will proceed
order by order. Up to third-order in moments, one can construct these conserved
operators just by directly promoting the conserved classical quantities \eqref{conserved_quantities} to operators: 
\begin{eqnarray}\label{r1op}
 \hat R_1=2 \hat H-\hat p,\\\label{r2op}
 \hat R_2={e^{2(\alpha+\hat\beta)}(\hat H-\hat p)}.
\end{eqnarray}
Up to the mentioned order, the expectation values of these operators can then be written in terms of moments
by performing an expansion around the expectation values of the basic variables
\begin{eqnarray}\label{expr1}
 \langle\hat R_1\rangle=R_1+\sum_{i+j=2}^3\frac{\partial^{i+j}R_1}{\partial\beta^i\partial p^j}\Delta(\beta^i p^j),\\
 \label{expr2}
 \langle\hat R_2\rangle=R_2+\sum_{i+j=2}^3\frac{\partial^{i+j}R_2}{\partial\beta^i\partial p^j}\Delta(\beta^i p^j).
\end{eqnarray}
Note that these expressions are valid for any ordering of the basic operators in the definitions \eqref{r1op}--\eqref{r2op}.
Any reordering of the operators would give a factor proportional to $\hbar^2$ in the above expansions,
which is of fourth-order and thus negligible at this level of approximation. Therefore, up to third order,
one can iteratively construct all the necessary constants of motion.

For instance, at second-order, the system is described by the two expectation values ($\beta$, $p$) and the
three moments $\Delta(\beta^ip^j)$ with $i+j\leq2$, namely
the correlation $\Delta(\beta p)$ and the two fluctuations $\Delta(\beta^2)$ and $\Delta(p^2)$. Therefore, in order to completely solve
the system, one needs to construct five constants of motion.
The expectation values \eqref{expr1}--\eqref{expr2} are two of them, and another three can be defined as,
\begin{eqnarray}
&&\langle (\hat R_1-R_1)^2 \rangle,\\
&&\langle (\hat R_1-R_1) (\hat R_2-R_2)+(\hat R_2-R_2)(\hat R_1-R_1) \rangle,\\
&&\langle (\hat R_2-R_2)^2 \rangle.\label{expr22}
\end{eqnarray}
Writing these expressions in terms of moments, as performed in \eqref{expr1}--\eqref{expr2},
and truncating the series at second-order,
one ends up with five constants of motion for five variables and the system is thus
completely determined.
Note that, for convenience, we have defined the conserved quantities as expectation
values of powers of differences, like $(\hat R_i-R_i)^2$, instead of expectation values
of powers, like $\hat R_i^2$, and with a symmetric ordering of $R_1$ and $R_2$.
In this way one automatically gets pure second-order real expressions
(there is no zeroth order contributions when expanding these expectation values).

The rationale within a third-order truncation scheme is exactly equivalent to the one applied at second order,
but with some more variables. In this
case the number of variables is nine: two expectation values ($\beta,p$), three second-order moments,
and four third-order moments (the two pure fluctuations $\Delta(\beta^3)$ and $\Delta(p^3)$, in combination
with the two high-order correlations $\Delta(\beta p^2)$, $\Delta(\beta^2 p)$).
In addition to the conserved quantities \eqref{expr1}--\eqref{expr22},
one can construct another four by computing the following expectation values
\begin{eqnarray}
&&\langle (\hat R_1-R_1)^3 \rangle,\\
&&\langle (\hat R_1-R_1)^2 (\hat R_2-R_2)+(\hat R_2-R_2)(\hat R_1-R_1)^2 \rangle,\\
&&\langle (\hat R_1-R_1) (\hat R_2-R_2)^2+(\hat R_2-R_2)^2(\hat R_1-R_1) \rangle,\\
&&\langle (\hat R_2-R_2)^3 \rangle.
\end{eqnarray}

From fourth order on, the situation gets more involved, as the ordering of the basic operators
in the definitions \eqref{r1op}--\eqref{r2op} comes into play. We find out that the
expectation value of the operator $\hat R_1$, with a completely symmetric ordering of basic operators,
\begin{equation}\label{r1weyl}
\hat{R}_1 = 2\hat{H}-\hat{p}\big|_{\text{Weyl}} 
\end{equation}
is conserved up to seventh-order in moments. Nonetheless, this is not the case for $\hat R_2$:
by considering a Weyl-ordered form for $\hat R_2$, its time derivative turns out not to be vanishing,
due to some terms proportional to $\hbar^2$. Still, we have been able to explicitly integrate
these non-vanishing terms and, by subtracting the result from the expectation value of the Weyl-ordered
operator version of $R_2$, construct the corresponding conserved quantity. This leads to the following
form for the second conserved operator,
\begin{widetext}
\begin{align}\label{r2weyl}
	\hat{R}_2 &= {e^{2(\alpha+\hat\beta)}(\hat H-\hat p)}+\hbar^2 {e^{2(\alpha +\hat \beta)}(\hat H-\hat p)^2\frac{15 \hat p \hat H^2+11\hat H^3+3\hat H \hat p^2-4 \hat p^3}{2 \hat H^4 \hat R_1^2}}\bigg|_{\text{Weyl}}.
\end{align}
\end{widetext}
In summary, up to seventh-order in moments, the two operators \eqref{r1weyl}--\eqref{r2weyl} are conserved.
Therefore, one can use these expressions to generate, at each order, all the necessary constants of motion
by computing expectation values of their products, as explained above for second- and third-order truncations.

\subsection{Quantum Kasner transition}
\label{sec_quantumtransition}

At this point, in order to obtain the quantum transition rules, one can proceed
systematically. One just needs to compute the expectation values of products of
the conserved operators \eqref{r1weyl}--\eqref{r2weyl}, and write them in terms of moments
by performing an expansion around the expectation values $\beta$ and $p$.
These constants of motion are then evaluated on each asymptotic Kasner regions,
by considering the behavior of different variables \eqref{quantum_Kasner_beta}--\eqref{quantum_Kasner_G_ij}.
The requirement of conservation of these quantities leads to a system of equations
that relates the parameters of the first Kasner epoch $(\overline c, \overline p, \overline{\Delta(\beta^{i}p^{j})})$
with those of the second Kasner regime $(\widetilde c, \widetilde p, \reallywidetilde{\Delta(\beta^{i}p^{j})})$.
This system of equations must then be solved to obtain the quantum transition rules.

For definiteness, we will complete this procedure with a fifth-order truncation,
by assuming that all moments $\Delta(\beta^i p^j)$ with $i+j>5=:N$ are negligible.
Up to this order, there are 20 variables and one needs to construct the same
amount of constants of motion.

Regarding the classical variables we find out that, on the one hand, making use of the conserved quantity
$\langle \hat{R}_1\rangle$, which takes the simple form $\langle\hat{R}_1\rangle= 2|p|-p$ at every order
in the Kasner regimes,
one can immediately obtain the transition rule for the momentum $p$,
\begin{align}
\label{law_p}
&\widetilde{p}=-\frac{1}{3}\overline p.
\end{align}
This expression keeps the same form as its classical counterpart \eqref{transition_p} and it is not affected by the moments.
On the other hand, the quantum back-reaction does affect the transition rule for the parameter $c$,
which encodes the asymptotic behavior of the shape-parameter $\beta$,
and it takes the form
\begin{align}
\label{law_beta} &\widetilde{c}=-3 \overline{c}-\frac{1}{2}\ln\bigg({\frac{2\overline p^2}{3}}\bigg)+\sum_{n=2}^{N}\frac{(-1)^n}{n\overline p^{n}}
\overline{\Delta({p^n})}.
\end{align}
However, note that only the initial relative pure fluctuations of the momentum $\overline{\Delta({p^n})}/ {\overline p}^n$
appear in this relation.

Concerning the moments, their transition rules are more involved, and they are explicitly
displayed in App. \ref{app:result}. Nevertheless, the general transition rule can be written
as follows,\footnote{There is just one exception to this general rule. As can be seen in App. \ref{app:result},
apart from the terms given in \eqref{betanpm}, the transition rule \eqref{exception} for the moment $\Delta(\beta^{3})$
contains an extra term proportional to $\hbar^2$.}
\begin{widetext}
\begin{align}\label{betanpm}
\reallywidetilde{\Delta(\beta^{n}p^{m})}=(-3)^{n-m}\overline{\Delta(\beta^{n}p^{m})}+\sum_{r=0}^{r_{max}}
\sum_{k=0}^{n-1}\sum_{l=l_{min}}^{l_{max}}
\frac{a_{nmklr}}{\overline{p}^{n+l+r-k}} \overline{\Delta(p^r)}\,\,\,\,
\overline{\Delta(\beta^k p^{n+m+l-k})},
\end{align}
\end{widetext}
with certain numerical coefficients $a_{nmklr}$. 
In fact, all explicit factors $p$ that appear in this expression can be absorbed by considering
relative moments with respect to $p$ (that is, $\Delta(\beta^n p^m)/p^m)$,
and by taking into account the transition rule \eqref{law_p},
\begin{widetext}
\begin{align}\label{betanpmrelative}
\frac{\reallywidetilde{\Delta(\beta^{n}p^{m})}}{\reallywidetilde p^m}=(-3)^{n}\frac{\overline{\Delta(\beta^{n}p^{m})}}{\overline p^m}+\sum_{r=0}^{r_{max}}
\sum_{k=0}^{n-1}
\sum_{l=l_{min}}^{l_{max}} b_{nmklr} \frac{\overline{\Delta(p^r)}}{\overline{p}^r}
\frac{\overline{\Delta(\beta^k p^{n+m+l-k})}}{\overline{p}^{n+m+l-k}},
\end{align}
\end{widetext}
with $b_{nmklr}=(-3)^ma_{nmklr}$. In this way, the transition rule for the
relative moments does not explicitly depend on the values of the asymptotic classical parameters
$p$ or $c$.

In the above expressions, the lower limit for $l$ is defined as $l_{min}=\max\{-r,k-(n+m)\}$.
This ensures that the right-hand sides of \eqref{betanpm} and \eqref{betanpmrelative} are exclusively
given by contributions of order higher or equal to $n+m$. Moreover, the upper limits of two of the sums, $r_{max}$ and $l_{max}$,
are an artifact of the truncation we are considering, and they just impose that the product of two
moments inside the sum to be, at most, of order $N$, that is, $n+m+l+r\leq N$. 
However, the upper limit in
the sum in $k$ is independent of the truncation, and thus physically meaningful. More precisely,
the largest possible value for $k$ is $n-1$, that is, lower that the index in $\beta$ of the final moment under consideration.
Therefore, this means that the asymptotic form of a moment $\reallywidetilde{\Delta(\beta^{n}p^{m})}$ after
the Kasner transition, depends on the initial value of the very same moment $\overline{\Delta(\beta^{n}p^{m})}$
and all moments $\overline{\Delta(\beta^k p^q)}$, with any $q\geq 0$
but with $0\leq k\leq n-1$. This translates into the fact that for a given moment,
the larger the index of the shape parameter $n$, the more complicate is the corresponding transition rule,
as more initial moments enter into play.

In particular, the relative pure fluctuations of the momentum $p$, that
is moments with $n=0$, have a very simple transition rule.
In fact, they have the same asymptotic value after and before the transition
\begin{equation}\label{ruledeltap}
 \frac{\reallywidetilde{\Delta (p^m)}}{\reallywidetilde p^m}= \frac{\overline{\Delta (p^m)}}{\overline{p}^m}.
\end{equation}
This is due to the fact that, at any order, the conserved quantity $\langle(\hat{R}_1-\langle{\hat{R}_1}\rangle)^m\rangle$
takes the simple form $\langle(\hat{R}_1-\langle{\hat{R}_1}\rangle)^m\rangle=(2\text{sign}(p)-1)^m\Delta(p^m)$
in the Kasner regime.

For $n=1$ the transition rule complicates a bit,
\begin{align}
\frac{\reallywidetilde{\Delta (\beta p^m)}}{\reallywidetilde p^m}&= -3 \frac{\overline{\Delta (\beta p^m)}}{\overline p^m}+\sum_{k=m+1}^{N}\frac{(-1)^{k-m}}{k-m}
\frac{\overline{\Delta (p^k)}}{\overline p^k}
\nonumber
\\[10pt]
\label{transition_Q(1m)}
&\qquad+
\sum_{k=2}^{N-m}\frac{(-1)^{k-1}}{k}
\frac{\overline{\Delta (p^m)}}{\overline p^m}
\frac{\overline{\Delta (p^k)}}{\overline p^k}
,
\end{align}
but, as can be seen, only moments $\overline{\Delta(\beta p^m)}$ and $\overline{\Delta(p^n)}$ appear on the right-hand side.
In order to check the transition rules for higher values of $n$, we refer the reader to App. \ref{app:result}.

Another important property of the transition rule \eqref{betanpmrelative} is that it is composed by linear ($r=0$) as well as
quadratic ($r\neq 0$) terms on initial moments. Nonetheless, concerning these quadratic combinations,
note that one of the components
is always given by pure fluctuations of the momentum $\overline{\Delta(p^r)}/\overline{p}^r$, which are conserved through the Kasner transition
\eqref{ruledeltap}.

In expressions \eqref{betanpm}--\eqref{betanpmrelative}, we have chosen to
explicitly write the contribution of the initial moment $\Delta(\beta^n p^m)$ on the right-hand
side, instead of including it in the sum. In this way, one can see that the transition produces two distinct effects
on the moments. 

On the one hand, the moment $\Delta(\beta^n p^m)$ is modified by a multiplicative factor $(-3)^{n-m}$.
Therefore, for moments with more weight on $\beta$ than on $p$, that is with $n>m$, the absolute value of
the moment is increased by this factor. On the contrary, for moments with a dominant contribution of $p$
($n<m$), this effect produces a decrease in their absolute value. Finally, moments with $n=m$ are not affected by this global factor.
All in all, this effect produces a squeezing of the state on the phase space by compressing it in the
$p$ direction, and stretching it in the $\beta$ direction.
The compression in the $p$ direction, given by the factor $(-3)^m$,
is clearly inherited from the transition rule for the momentum \eqref{law_p}.
This is why the relative moments $\Delta(\beta^n p^m)/p^m$ do not suffer such compression in $p$.
The effect in the $\beta$ direction is also related to the factor $-3$ that appears in 
the transition rule \eqref{law_beta} for the parameter $c$ that defines the asymptotic behavior of $\beta$.
But since, in this case, there are other moments involved, it is not straightforward
to absorb the factor $(-3)^n$ in the definition of certain relative moments.

On the other hand, the moment $\Delta(\beta^n p^m)$ is modified by the presence of other moments
$\Delta(\beta^k p^q)$, with $0\leq k\leq n-1$. This effect, which couples the final value of a given
moment with the initial value of other moments, has not a definite sign, since the numerical
factors $a_{nmklr}$ can be either positive or negative. Therefore, the interpretation of this
effect is not so clear as the previous one, and one has to check individual moments (and particular
states) to see how each moment is affected.

For instance, if we consider an initial Gaussian state in the momentum $p$,
these effects can easily been observed. Its corresponding moments, \textit{i.e.}, the asymptotic
moments before the transition, are given as follows
\begin{equation}
\label{moments_gaussian}
\overline{\Delta(\beta^{2n}p^{2m})}= 2^{-2(n+m)}
\hbar^{2n}\sigma^{2(m-n)}\frac{(2n)!(2m)!}
{n!m!},
\end{equation}
for $\forall n,m\in\mathbb{N}$, and vanishing otherwise. These moments will evolve and after the transition,
they will transform according to \eqref{betanpm}.
For the sake of clarity, let us focus on how the second-order moments are modified
\begin{align}
	\label{G11_gaussian}
	&\reallywidetilde{\Delta(\beta p)}
	=\frac{\overline{\Delta(p^2)}}
	{3\overline{p}}
	+\frac{\overline{\Delta(p^4)}}
	{9\overline{p}^3},
	\\[10pt]
	\label{G02_gaussian}
	&\reallywidetilde{\Delta(p^2)}=
	\frac{1}{9}\overline{\Delta(p^2)},
	\\[10pt]
	\label{G20_gaussian}
	&\reallywidetilde{\Delta(\beta^2)}
	=
	9\overline{\Delta(\beta^2)}
	+\frac{\overline{\Delta(p^2)}}{\overline{p}^2}
	\Bigg(1-\frac{\overline{\Delta(p^2)}}{4\overline{p}^2}
	\Bigg)
	+\frac{11}{12}\frac{\overline{\Delta(p^4)}}{\overline{p}^4},
\end{align}
$\overline{p}$ being the value of $p$ in the initial Kasner regime.
First of all, from \eqref{G11_gaussian} we observe that $\reallywidetilde{\Delta(\beta p)}>0$,
which means that the transition generates a positive correlation between $\beta$ and $p$.
Thus, since this correlation is vanishing for any Gaussian state, it is clear that the final state does not belong to this class. Moreover, from
relation \eqref{G02_gaussian} one can see that the initial state is compressed in the $p$ direction.
However, from the transition rule \eqref{G20_gaussian} it is not straightforward to deduce whether
the initial state is stretched or compressed in the $\beta$ direction. Nevertheless, by considering
the explicit form (\ref{moments_gaussian}) for the initial moments,
it can be shown that,
\begin{align}
\label{G20_gaussian_explicit}
&\frac{\reallywidetilde{\Delta(\beta^2)}}
{\overline{\Delta(\beta^2)}}
=
\frac{\frac{5 \sigma ^6}{4 \overline{p}^4}+\frac{\sigma ^4}{\overline{p}^2}}{\hbar^2}+9,
\end{align}
which is clearly larger than one.
Therefore, an initial Gaussian state is certainly stretched in the $\beta$ direction.

Finally, it is also interesting to evaluate the Heisenberg uncertainty principle which,
for the final state takes the form,
\begin{align}
&
 \reallywidetilde{\Delta(p^2)}\,
\reallywidetilde{\Delta(\beta^2)}
-\big(\reallywidetilde{\Delta(\beta p)}\big)^2
-
\frac{\hbar^2}{4}
=\frac{\sigma ^6 \left(\overline{p}^2-\sigma ^2\right)}{144 \overline{p}^6}\geq 0.
\nonumber
\label{Heisenberg-gaussian}
\end{align}
In order to be satisfied, the relative initial fluctuation of the momentum
should be small, $\sigma/p \leq 1$, which
is consistent with our assumption of peaked semiclassical states.

\subsection{Analysis of the transition dynamics for a Gaussian initial state}
\label{sec_numerics}

In this section we are interested in examining in detail the evolution of the state
and, more specifically, the transition dynamics. In order to develop this study,
we will need to perform a numerical analysis. For this purpose, let us consider
an initial unsquezeed Gaussian state in the initial Kasner epoch, characterized
by a Gaussian width $\sigma=\sqrt{\hbar}$. The moments for such a state take the form
given in equation \eqref{moments_gaussian}.
In addition, for the rest of the variables,
we choose as initial conditions $p=1$, $\beta=100$ and $\alpha=100$. In this way
we ensure that the potential of the Hamiltonian is negligible and thus the
system begins in a Kasner regime.

As expected, the numerical study shows that, evolving the system towards lower values
of $\alpha$, the system follows the Kasner dynamics until it undergoes a transition,
which happens near $\alpha=0$. This transition takes place in a very small period of
time and then the system reaches its final equilibrium value --the final Kasner regime-- very quickly.

\begin{figure}
	\centering
	\includegraphics[width=0.7\linewidth]{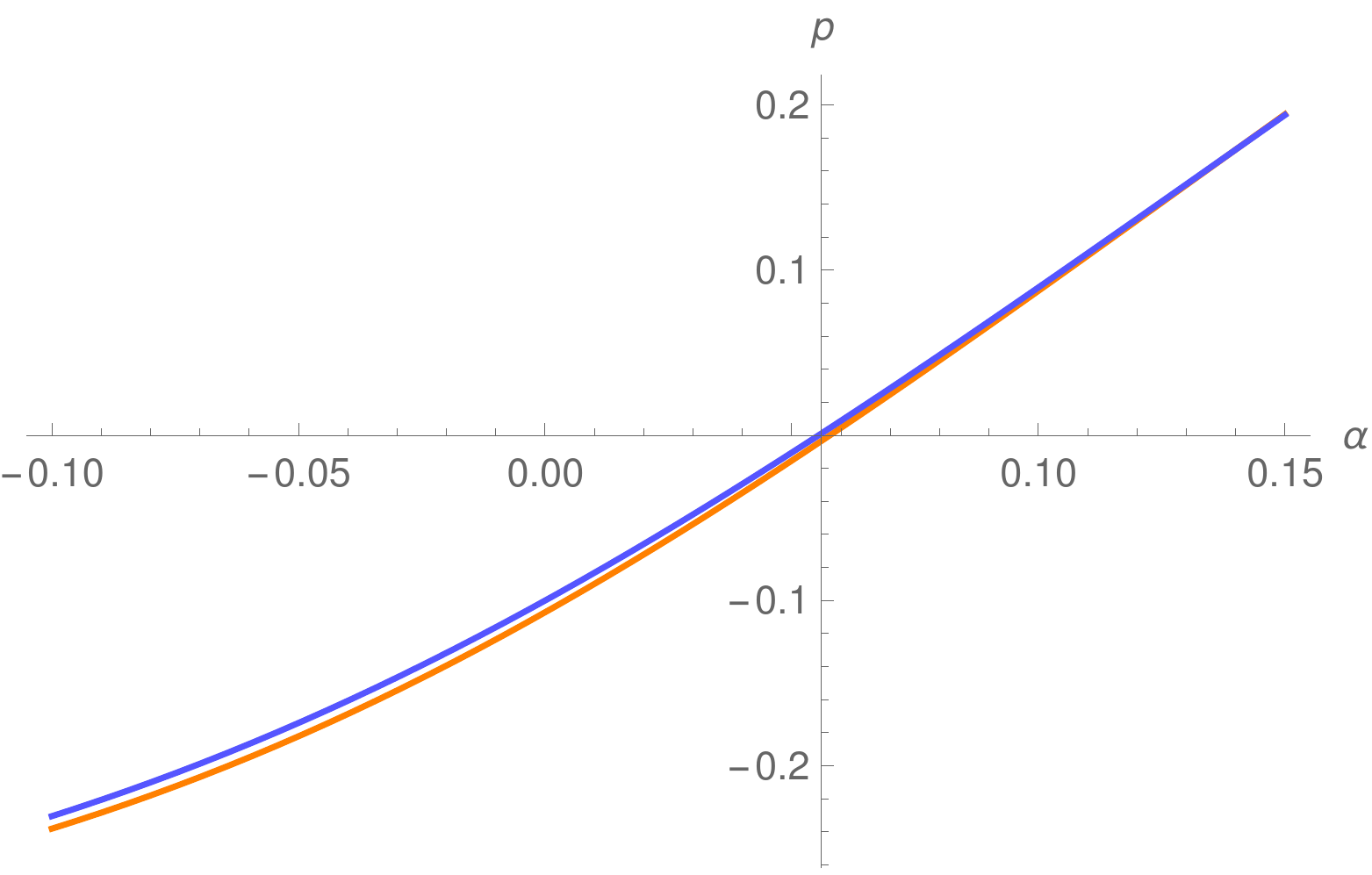}
	\caption{Comparison of classical and quantum evolution of the momentum $p$ during the transition. The orange line represents the classical evolution and the blue one the quantum evolution.}
	\label{fig:pgaussian}
\end{figure}

\begin{figure}
	\centering
	\includegraphics[width=0.7\linewidth]{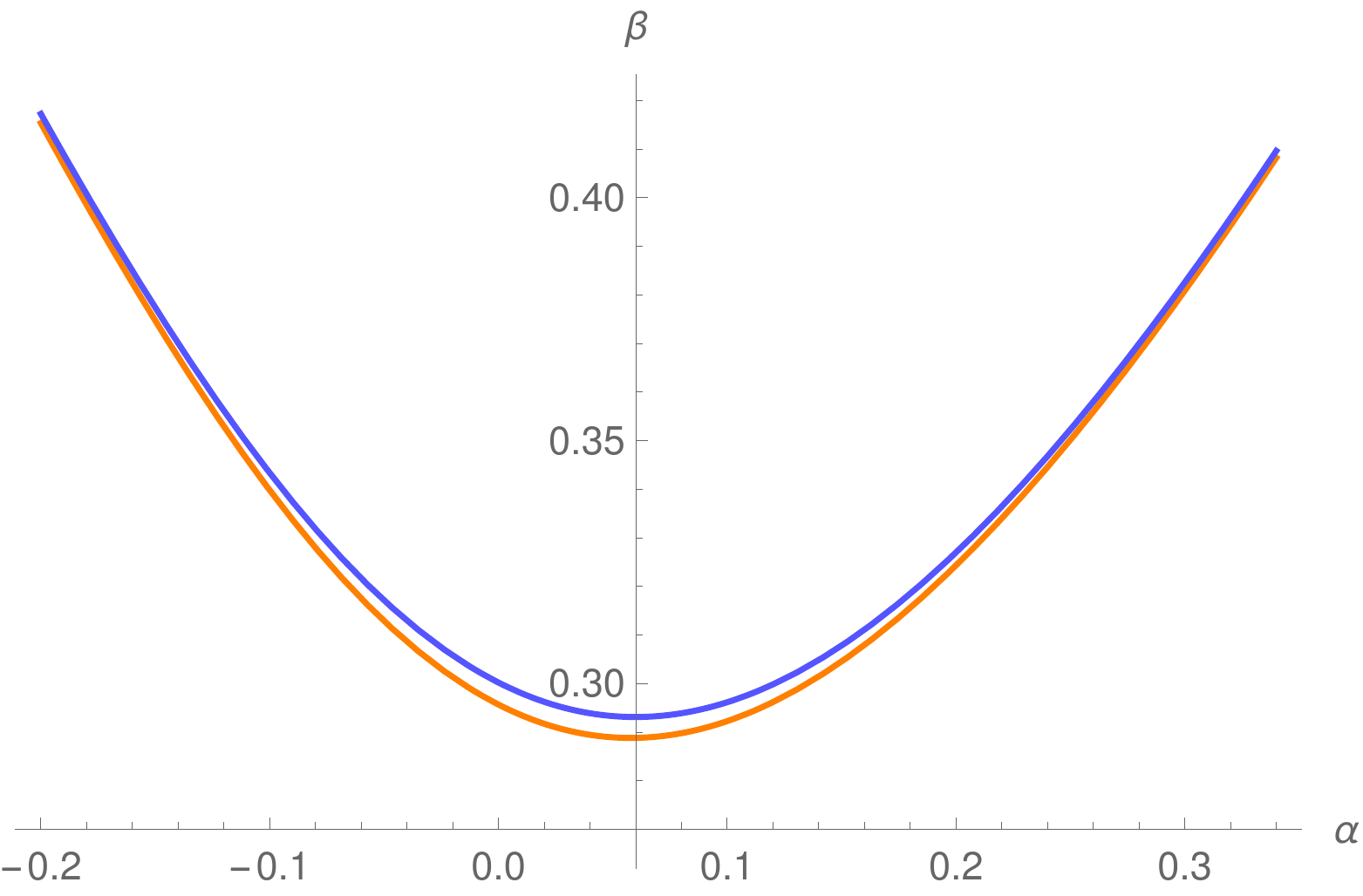}
	\caption{Comparison of classical and quantum evolution of the shape-parameter $\beta$ during the transition. The orange line represents the classical evolution and the blue one the quantum evolution.}
	\label{fig:betagaussian}
\end{figure}

On the one hand, concerning the evolution of the expectation values
$p$ and $\beta$, we conclude that they remain almost identical as in the classical case, that is, as depicted in Figs. \ref{fig:beta} and \ref{fig:p}.
In fact, only in a very close zoom into the transition (near $\alpha=0$) we can appreciate a slight modification in their evolution,
which is shown in Figs. \ref{fig:pgaussian} and \ref{fig:betagaussian}. In particular, from these figures one can see that quantum effects
slightly increase the value of $p$ and $\beta$ around the transition.

\begin{figure}[]
	\centering
	\includegraphics[width=0.7\linewidth]{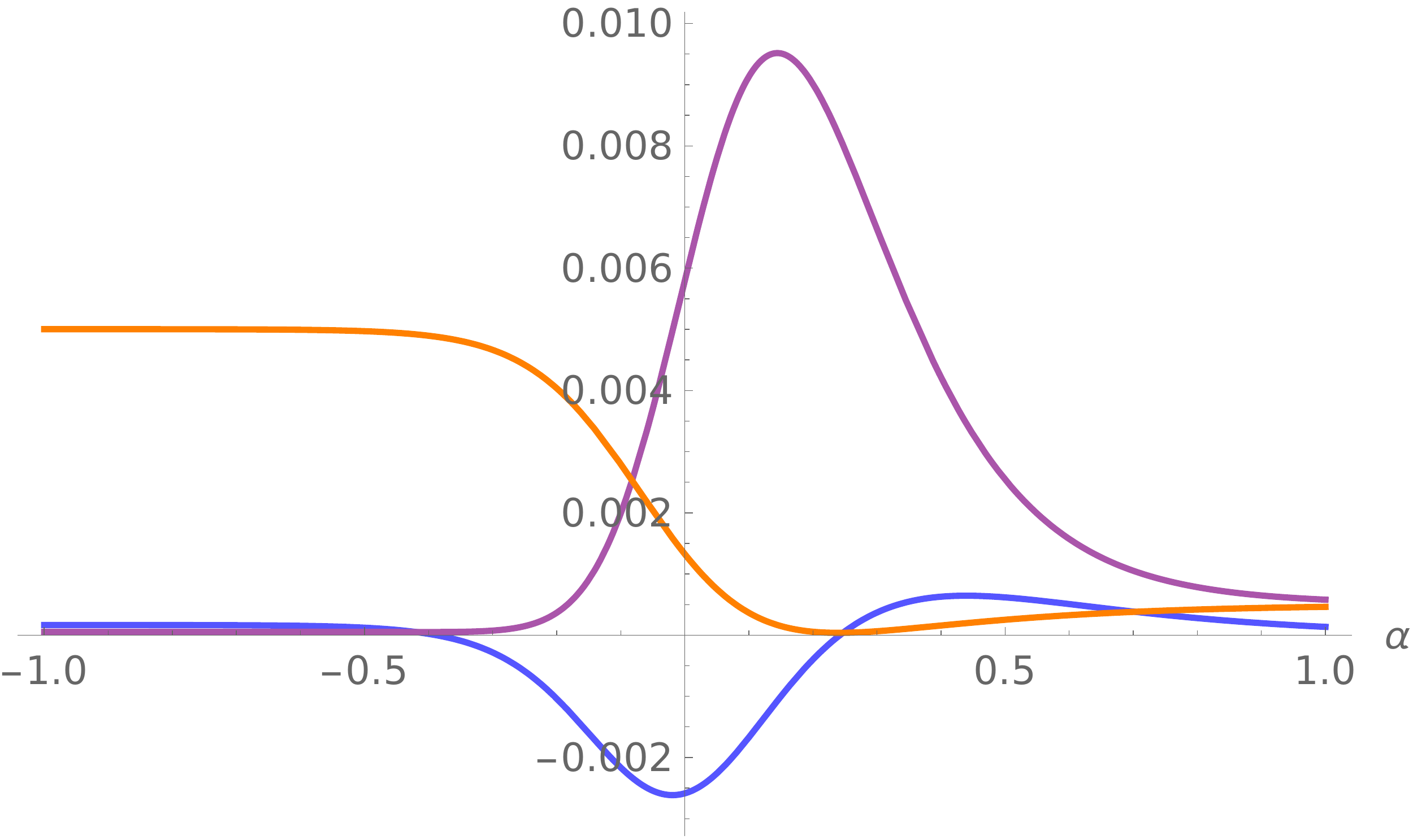}
	\caption{Evolution of the second-order moments. $\Delta(\beta p)$ is depicted in blue, $\Delta(p^2)$ in purple and $\Delta(\beta^2)$ in orange.
	During the transition a small positive correlation is generated, the fluctuation of the shape-parameter is amplified, and the fluctuation
	of the moment is diminished.}
	\label{fig:second_order}
\end{figure}

\begin{figure}[]
	\begin{center}
		\includegraphics[width=0.7\linewidth]{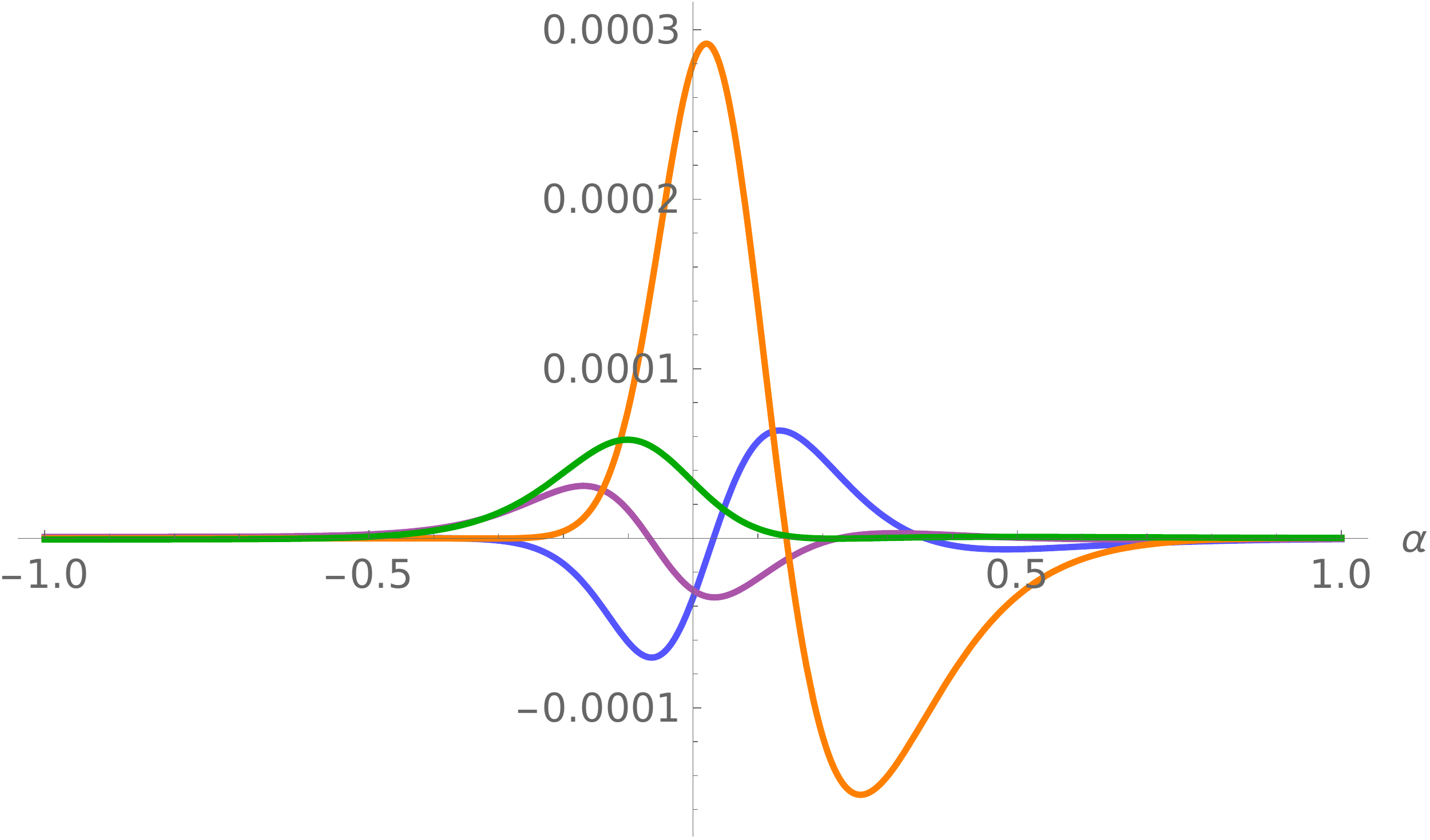}
		\includegraphics[width=0.7\linewidth]{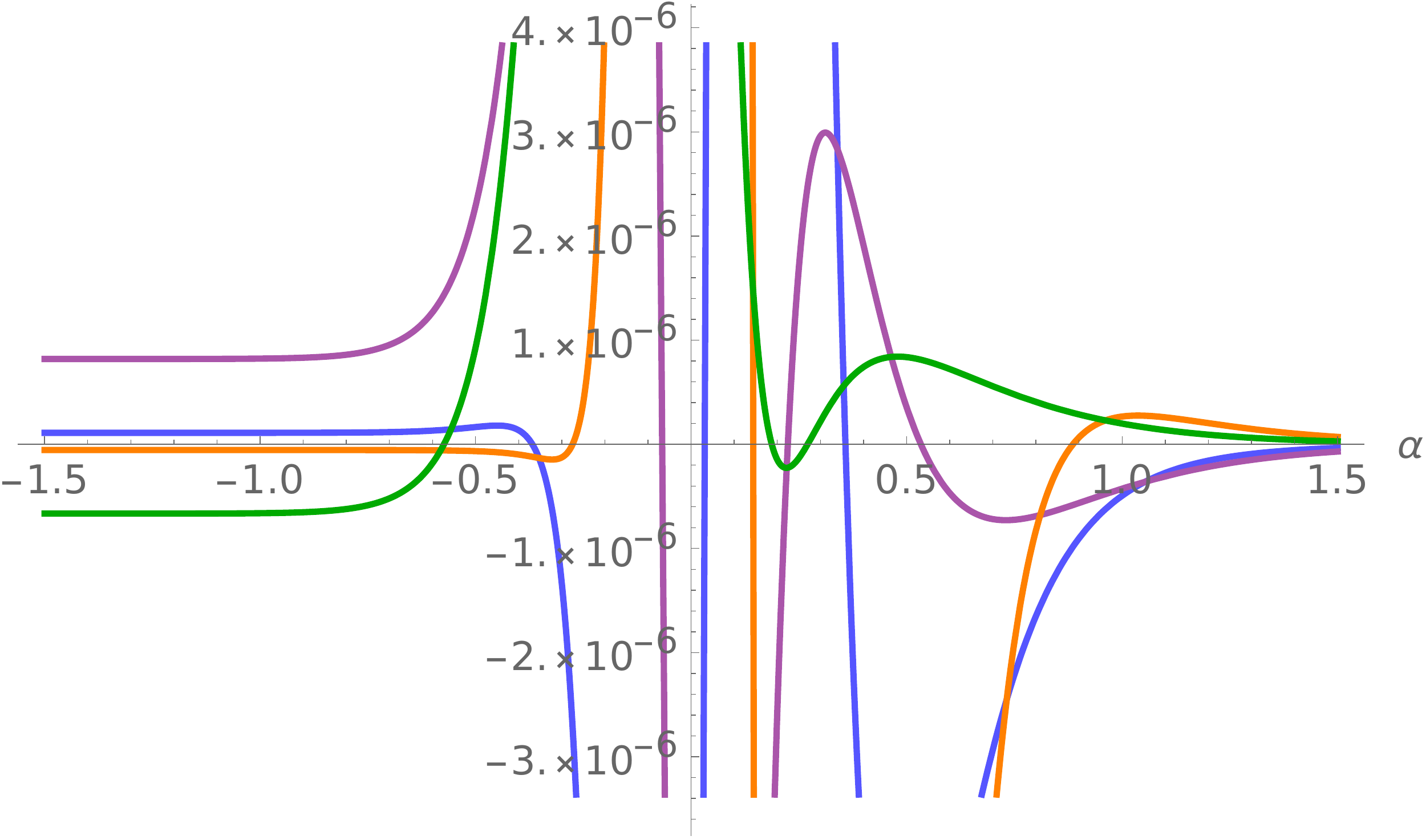}
		\caption{Evolution of the third-order moments, with the whole range of the amplitude in the first plot and with a zoom in the second. $\Delta(p^2\beta)$ is depicted in blue, $\Delta(\beta^2p)$ in purple, $\Delta(p^3)$ in orange and $\Delta(\beta^3)$ in green.}
		\label{fig:third_order}
	\end{center}
\end{figure}

\begin{figure}[]
	\begin{center}
		\includegraphics[width=0.7\linewidth]{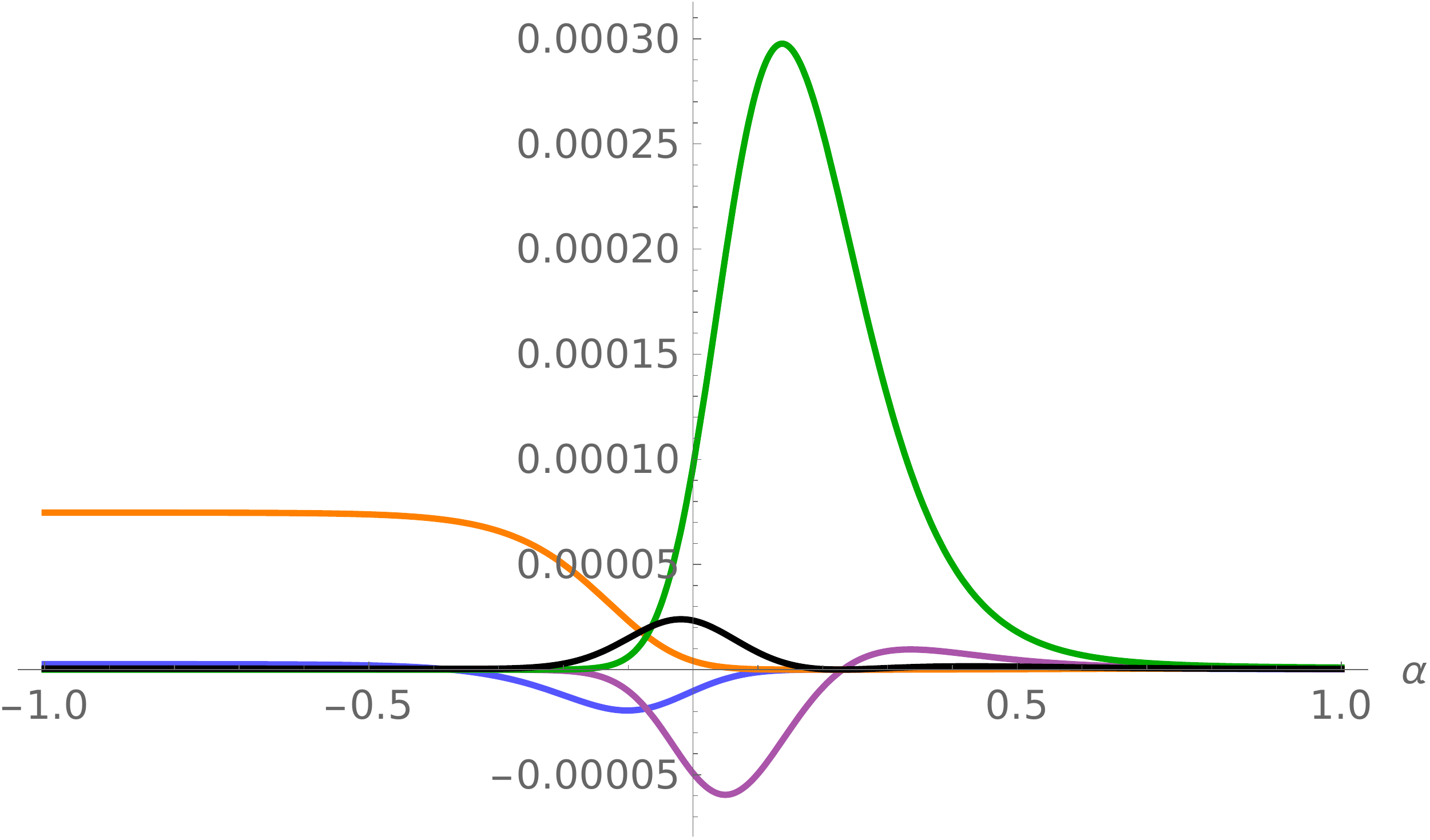}
		\includegraphics[width=0.7\linewidth]{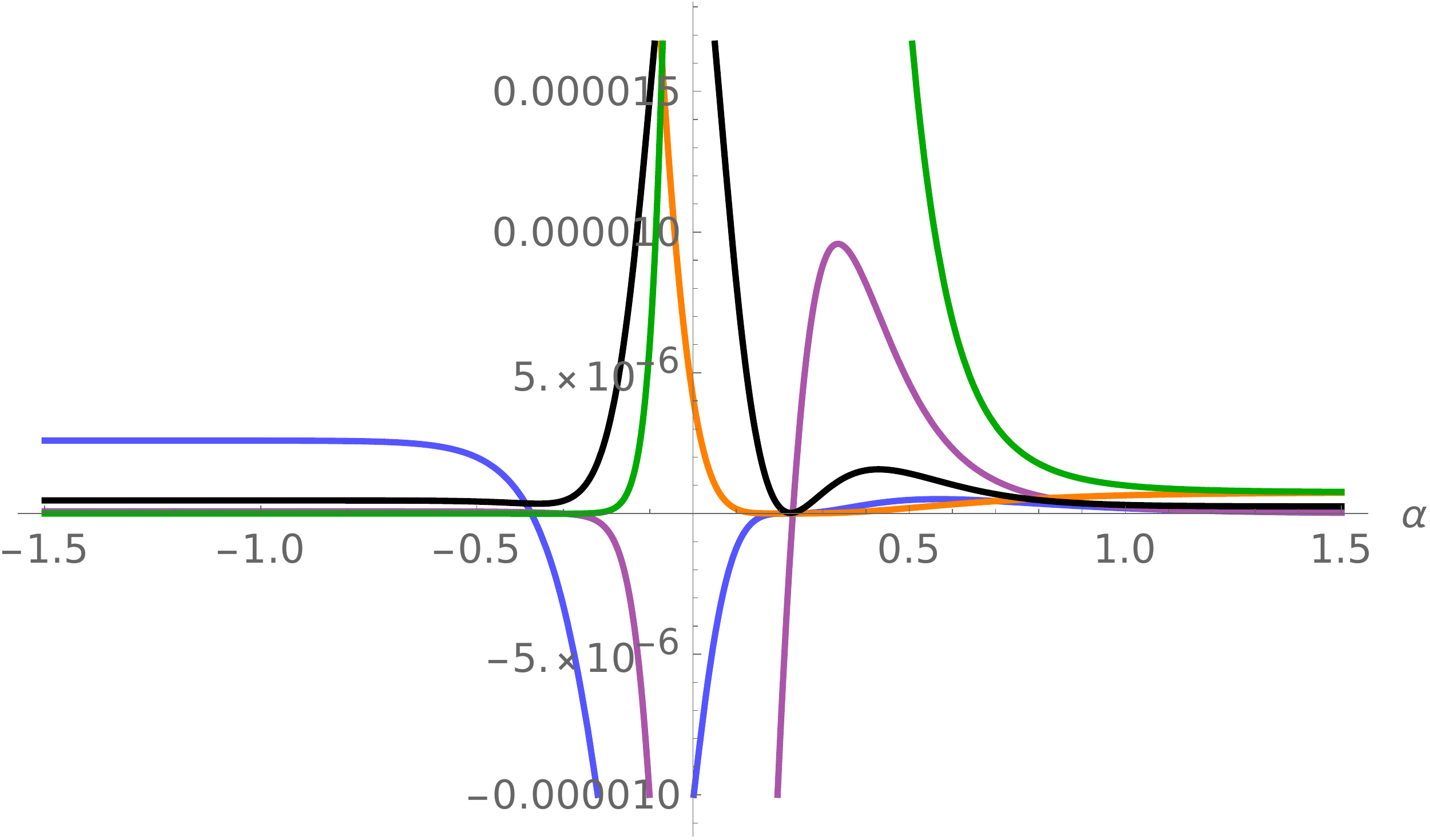}
		\caption{Evolution of the fourth-order moments, with the whole range of the amplitude in the first plot and with a zoom in the second. $\Delta(\beta^3p)$ is depicted in blue, $\Delta(\beta p^3)$ in purple, $\Delta(\beta^4)$ in orange, $\Delta(p^4)$ in green and $\Delta(\beta^2p^2)$ in black.}
		\label{fig:fourth_order}
	\end{center}
\end{figure}

\begin{figure}
	\begin{center}
		\includegraphics[width=0.7\linewidth]{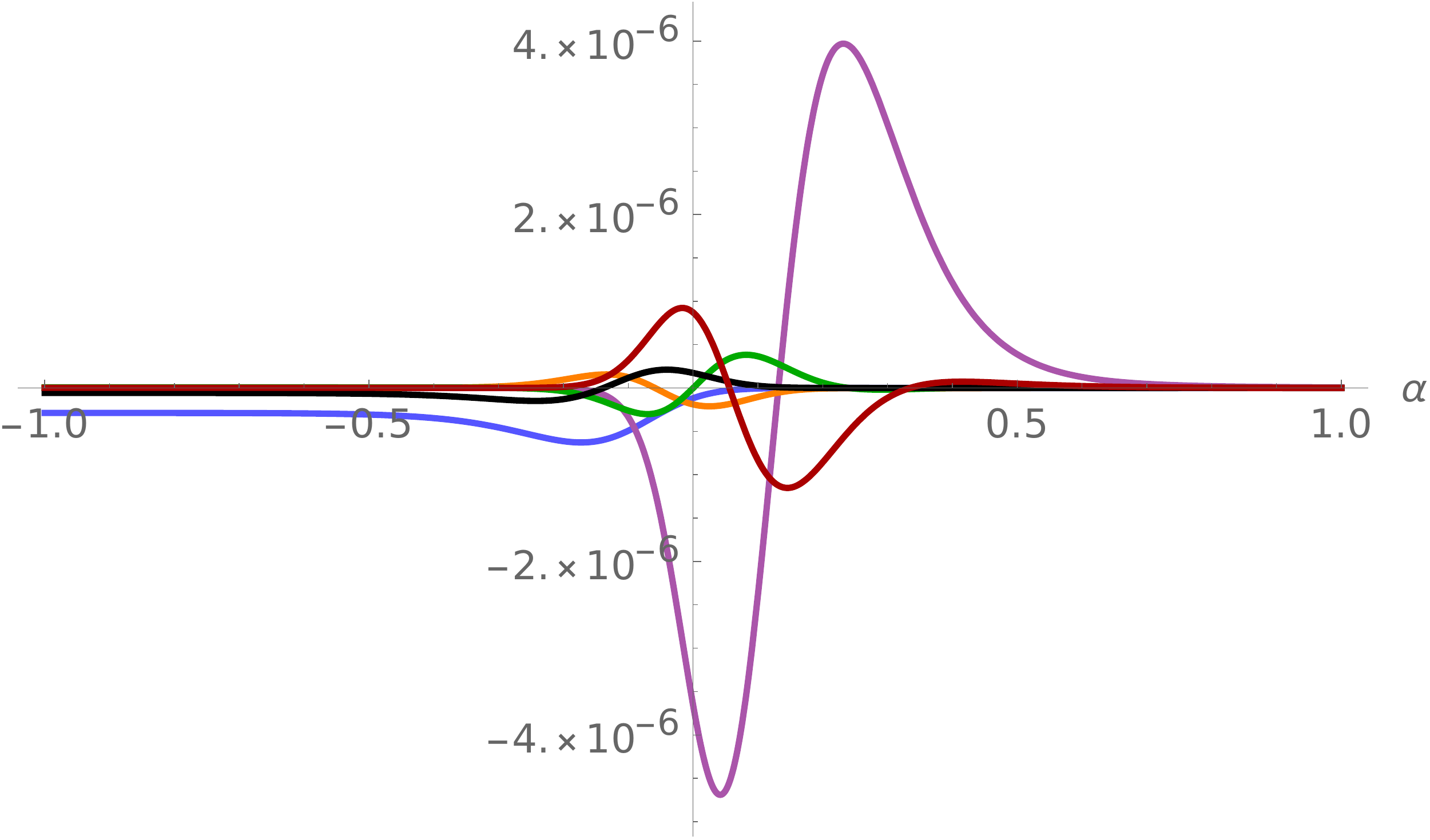}
		\includegraphics[width=0.7\linewidth]{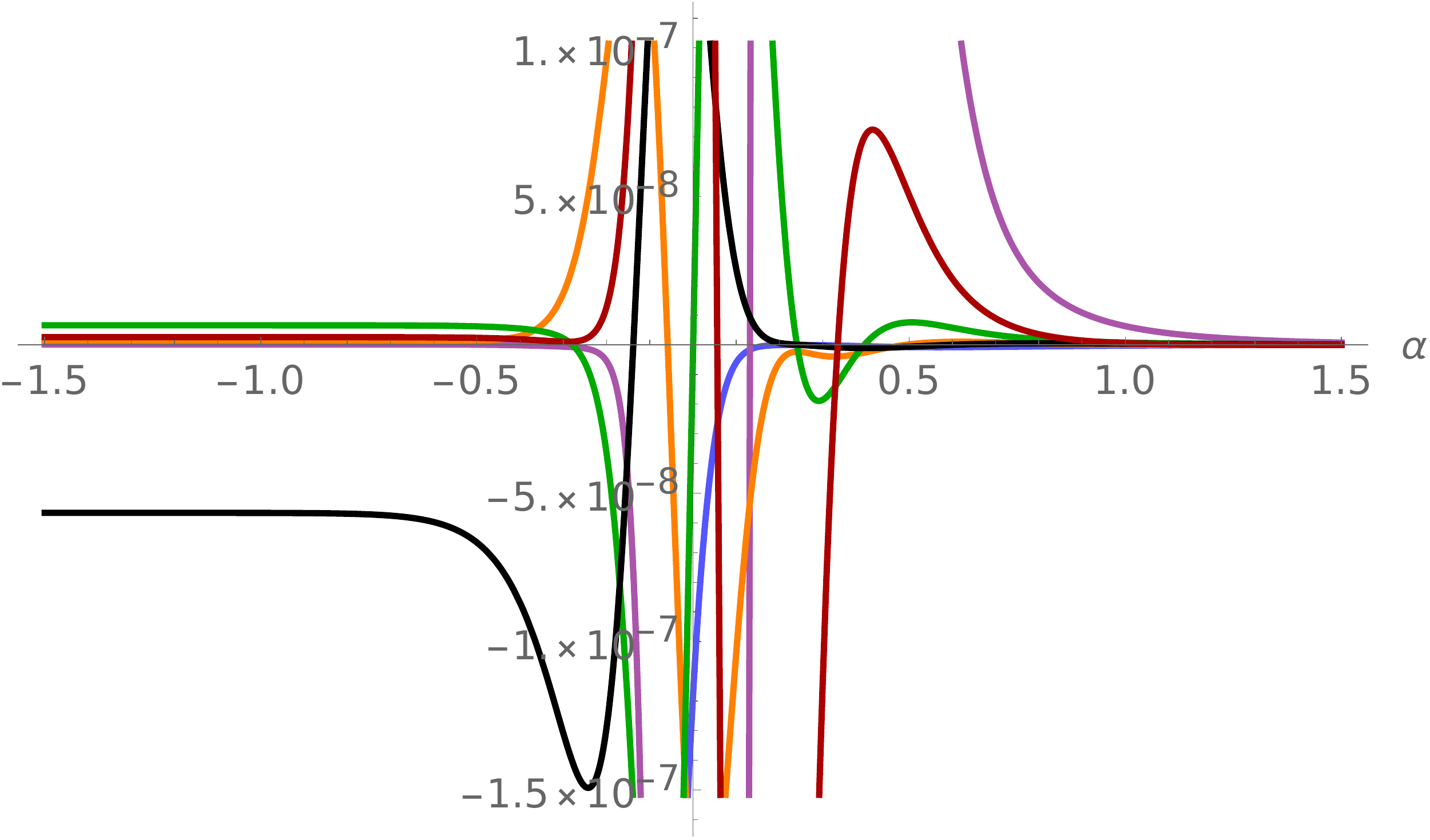}
		\caption{Evolution of the fifth-order moments, with the whole range of the amplitude in the first plot and with a zoom in the second. $\Delta(\beta^5)$ is depicted in blue, $\Delta(p^5)$ in purple, $\Delta(\beta^3p^2)$ in orange, $\Delta(\beta^2p^3)$ in green, $\Delta(\beta^4p)$ in black and $\Delta(\beta p^4)$ in red.}
		\label{fig:fifth_order}
	\end{center}
\end{figure}

On the other hand, in Figs. \ref{fig:second_order}--\ref{fig:fifth_order} the evolution of the moments has been plotted. As it can be seen, during the initial Kasner regime they hold a constant value, at certain point they begin to perform strong oscillations but then they quickly relax to another constant value,
which characterizes the coherent state during the final Kasner regime. It is particularly interesting to observe that the moments that were
vanishing in the initial Kasner epoch --due to the choice of the Gaussian state-- are not vanishing in the final one.
This is indeed an expected outcome according to the transition laws that we have discussed in the previous section.
In particular, this means that even when we choose an initial state where there are no correlations between $\beta$ and $p$,
the transition generates correlations for the final state. 

\begin{figure}
	\centering
	\includegraphics[width=0.7\linewidth]{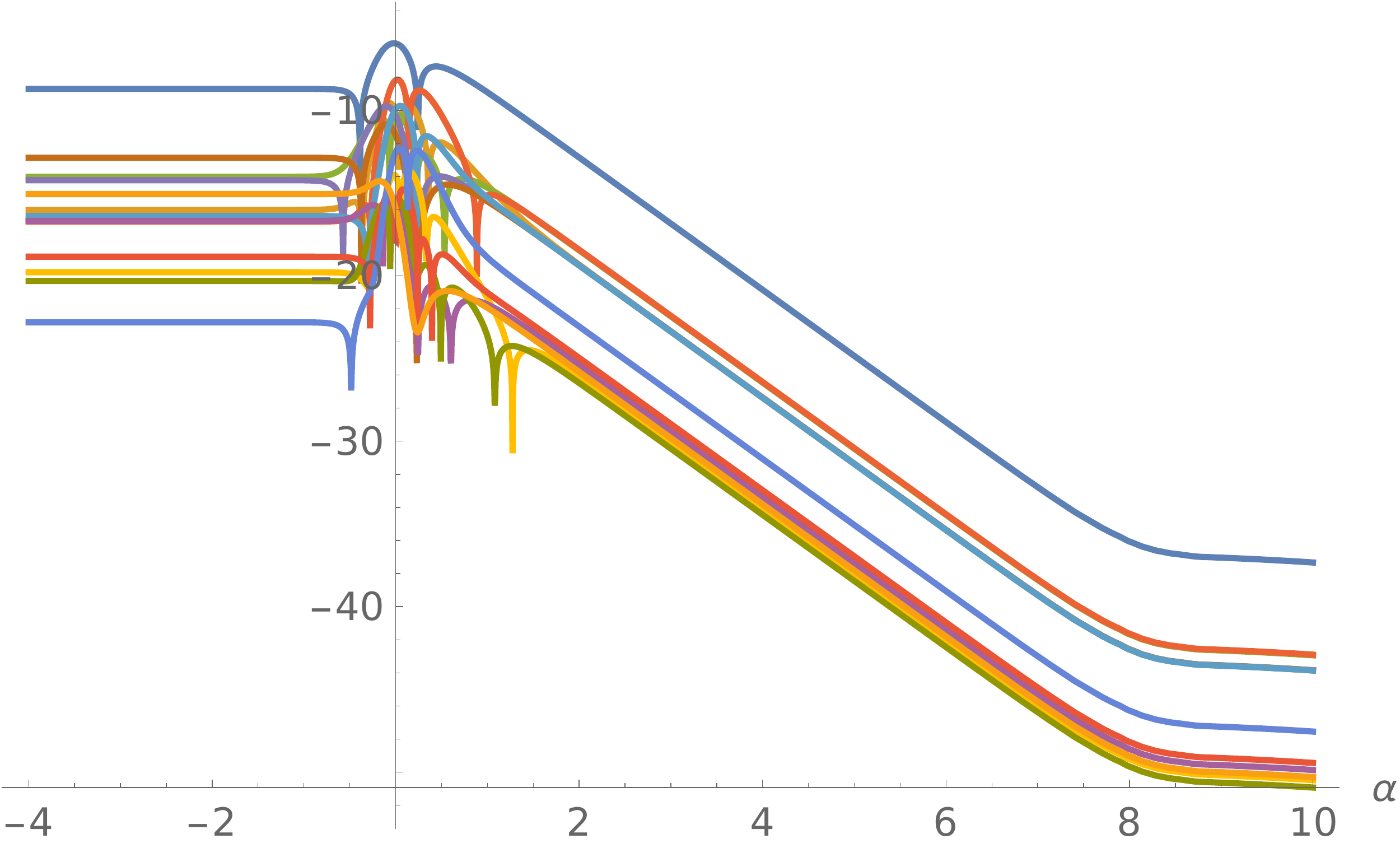}
	\caption{Logarithmic evolution of the initial vanishing moments with respect to $\alpha$ as they approach the transition.}
	\label{fig:activation}
\end{figure}

In fact, all the initial vanishing moments are activated in a similar way: as they approach the transition they experience an excitation
and start to grow exponentially, until they begin to oscillate. This behavior is shown in Fig. \ref{fig:activation}, where one can notice
that all of them grow with the same slope in a logarithmic plot. This slope is related to the potential of the Hamiltonian
and is found to be equal to $-4$. That is, between the exact Kasner regime and the transition,
the moments begin to feel the presence of a nonvanishing potential and their evolution is driven
by $\Delta(\beta^np^m)\propto e^{-4\alpha}$. Furthermore, we see that the lower the order of a given moment, the sooner it is activated, which is in agreement with
the semiclassical hierarchy of moments we are assuming. The same applies to the beginning of the oscillation
period: lower-order moments begin their oscillation regime before higher-order ones.


The precise dynamics of the moments is quite complicate but, if we look into the transition in more detail,
there are some general features that one can see. For instance, we observe that lower-order moments
tend to oscillate with bigger amplitudes as higher-order ones, as can be seen in Figs. \ref{fig:second_order}--\ref{fig:fifth_order}.
Moreover, at each given order, the pure fluctuations of $p$ have the biggest amplitudes. And the higher the index of $p$
the earlier the moments reach their final equilibrium value.

\begin{figure}[]
	\centering
	\includegraphics[width=0.6\linewidth]{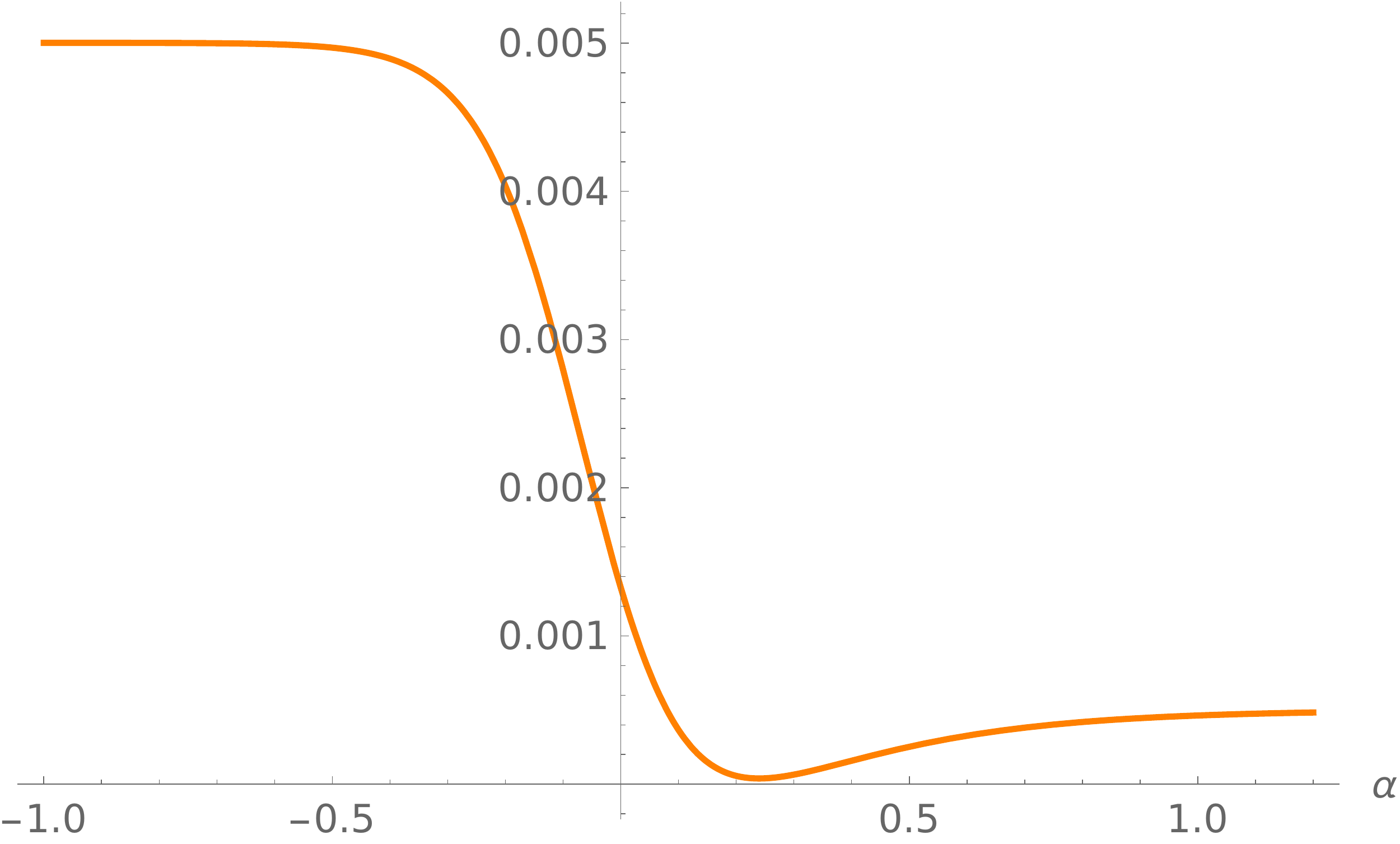}
	\includegraphics[width=0.6\linewidth]{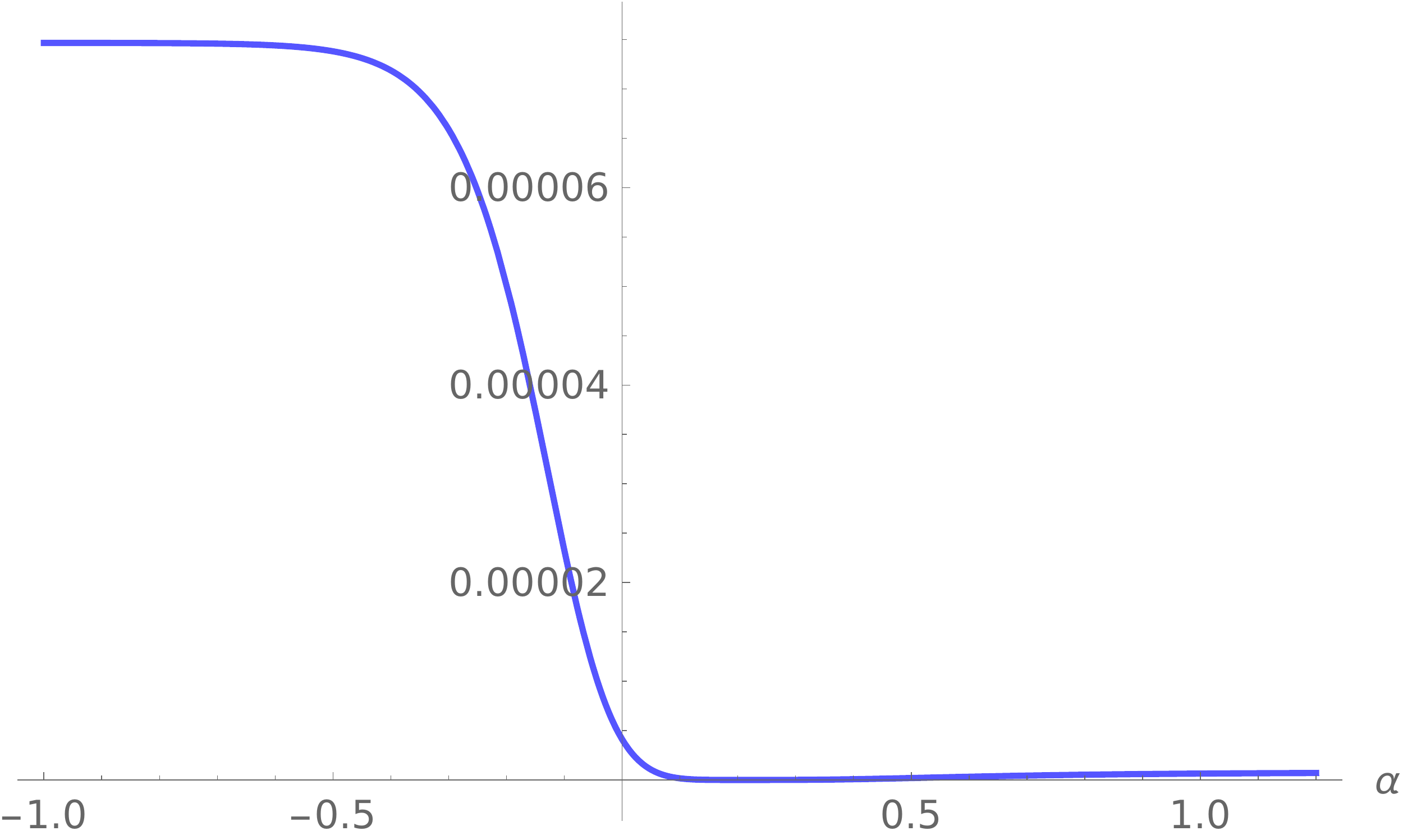}
	\caption{Evolution of $\Delta(\beta^2)$ and $\Delta(\beta^4)$ during the transition, in the first and second plot respectively.}
	\label{fig:G02-G04}

\end{figure}

Furthermore, in general, it can be seen that moments with at least one odd index, \textit{i.e.}, the initially vanishing ones, present more oscillations.
In fact, as shown in Fig. \ref{fig:G02-G04}, pure fluctuations of the shape-parameter $\Delta(\beta^n)$ with an even index $n$ do not oscillate per se;
they just have a local minimum in the transition and then grow up to their final value.

\section{Discussion} \label{sec:discuss}

In this paper, we have considered a quantum locally rotationally symmetric Bianchi II
model in order to analyze the transition between Kasner epochs. The behavior of the classical variables through
this transition is well-known, but there is an open discussion about how this behavior might
be modified when considering quantum effects that are expected to be relevant close to the singularity.

In order to describe the dynamics of this system,
we have considered Misner variables, which
are given by the (logarithm of the) spatial volume and two anisotropic shape-parameters.
In the locally rotationally symmetric
case under consideration, one of the shape-parameters is vanishing. In addition,
the Hamiltonian constraint is deparametrized by choosing the (logarithm of the) spatial volume as the
internal time, which leaves the system with one physical degree of freedom $(\beta, p)$,
$\beta$ being the nonvanishing shape-parameter and $p$ its conjugate momentum.

Once performed the quantization of the system, the whole information of the
quantum state is encoded in the infinite set of moments \eqref{moments}. Our analysis then
shows that, for peaked semiclassical states, the classical qualitative picture
is maintained in the quantum regime: there are two asymptotic Kasner regions,
connected by a quick transition. In particular, during the Kasner epochs,
the state evolves coherently, as $\langle \hat\beta\rangle$ and $\langle \hat p\rangle$
follow a classical trajectory on the phase space, whereas all the fluctuations and higher-order moments are kept
constant throughout evolution.

By considering a fifth-order truncation in the moments, which corresponds to neglecting
terms of an order $\hbar^3$ and higher, we have been able to obtain the explicit
analytic transition rules, that relate the parametrization of the state before and
after the transition. These rules, schematically given by equations \eqref{law_p}--\eqref{betanpm}
and explicitly displayed in App. \ref{app:result}, comprise the most relevant result
of the present paper.

The transition rule for the momentum $p$ \eqref{law_p} remains unchanged
with respect to its classical counterpart.
Nonetheless the transition rule for the parameter $c$ \eqref{law_beta}, which characterizes the asymptotic
behavior of the shape-parameter $\beta$, is slightly modified by quantum back-reaction
effects. Interestingly, this back-reaction is completely encoded in the relative pure fluctuations
of the momentum $\Delta(p^n)/p^n$.

Concerning the moments, their transition rule \eqref{betanpm} is much more involved,
but one can notice certain general properties. In particular, the asymptotic form of
a given moment $\Delta(\beta^np^m)$ after the transition is given by certain combination of
the initial value of all moments $\Delta(\beta^kp^q)$ with $k\leq n$ and any $q$.
Therefore, the complexity of the transition rule is largely governed by
the index of the shape-parameter $n$. Furthermore, one can see
that the transition introduces two distinct deformations on the state.
On the one hand, the initial value of the moment is multiplied by certain
coefficient, which produces a squeezing of the state on the phase space
by compressing it in the $p$ direction and stretching it in the $\beta$ direction. On the other hand, the back-reaction of other moments enters
as an additive deformation in the transition rule. The physical 
interpretation of this second effect is though not so clear.

Finally,
in order to understand more precisely the dynamics during the transition,
in the last section, we have performed some numerical simulations for
an initial Gaussian state. The state is deformed throughout the transition,
as expected from the transition rules, and the initial vanishing
moments are excited. In particular the transition generates a positive
correlation of the system. In addition, we observe that the constant
asymptotic values of the moments are connected through quick and strong
oscillations.

Let us remark that this analysis has been carried out for a particular Kasner transition.
We have developed then a first approach to understand the prevalence and
changes of Kasner transitions, which play a key role in the chaotic oscillatory BKL behavior,
under quantum effects. A further study to a general case should be analyzed
in the future to have a full understanding of the quantum survival of the BKL transitions.

\section*{Acknowledgments}
AA-S is supported by the ERC Advanced Grant No. 740209.
SFU acknowledges financial
support from an FPU fellowship of the Spanish Ministry of Universities.
This work is funded by Projects FIS2017-85076-P and FIS2017-86497-C2-2-P (MINECO/AEI/FEDER, UE),
and by Basque Government Grant No.~IT956-16.

\appendix

\section{Quantum transition rules} \label{app:result}

In this appendix we display the complete explicit transition rules
for the $20$ parameters that describe the Kasner regime of the system up to fifth-order
in moments, that is, $p$, $c$, and all moments $\Delta(\beta^n p^m)$ with $n+m\leq 5$.
In order to remove all explicit dependence on $p$ from the expressions,
we provide the transition rules for the relative moments $Q^{nm}:=\Delta(\beta^np^m)/p^m$.
As in the main text, the objects before the transition are denoted with a bar $(\overline p, \overline c, \overline Q^{nm})$,
whereas after the transition they are denoted with a tilde  $(\widetilde p, \widetilde c, \widetilde Q^{nm})$:
\begin{widetext}
	\begin{align}
	\widetilde{p}&=-\frac{1}{3}p,
	\\
	\widetilde{c}&=-3c-\frac{1}{2}\ln\bigg(
	{\frac{2p^2}{3}}\bigg)+\frac{1}{2}
	\overline{Q}^{02}-\frac{1}{3}
	\overline{Q}^{03}
	+\frac{1}{4}\overline{Q}^{04}-\frac{1}{5}
	\overline{Q}^{05},
\end{align}
\begin{align}
	\widetilde{Q}^{11}&=-3\overline{Q}^{11}-\overline{Q}^{02}+\frac{1}{2}\overline{Q}^{03}-\frac{1}{3}\overline{Q}^{04}+\frac{1}{4}\overline{Q}^{05},
	\\
	\nonumber
	\widetilde{Q}^{20}&=\overline{Q}^{02}+6\overline{Q}^{11}+9\overline{Q}^{20}-\overline{Q}^{03}-3\overline{Q}^{12}
		+2\overline{Q}^{13}
	-\frac{1}{4}\big(\overline{Q}^{02}\big)^2+\frac{11}{12}\overline{Q}^{04}+ \frac{1}{3}
	\overline{Q}^{02}\overline{Q}^{03}
	\\
	&\quad-\frac{5}{6}\overline{Q}^{05}
-\frac{3}{2} \overline{Q}^{14},
	\\
	\widetilde{Q}^{02}&=\overline{Q}^{02},
	\\
	\nonumber
	\widetilde{Q}^{30}&= -\frac{9}{4p^2}\hbar^2
	-27\overline{Q}^{30}-27\overline{Q}^{21}-9\overline{Q}^{12}-\overline{Q}^{03}
	-\frac{3}{2}\big(\overline{Q}^{02}\big)^2+\frac{3}{2}\overline{Q}^{04}+9\overline{Q}^{13}
	\\
	\nonumber
	&\quad
	+\frac{27}{2}\overline{Q}^{22}
	-9\overline{Q}^{02}\overline{Q}^{11}
	-\frac{27}{2}\overline{Q}^{02}\overline{Q}^{20}
	-\frac{7}{4}\overline{Q}^{05}-\frac{33}{4}\overline{Q}^{14}
	-9\overline{Q}^{23}
	\\
	&\quad
	+6\overline{Q}^{03}\overline{Q}^{11}
	+9\overline{Q}^{03}\overline{Q}^{20}
	+\frac{5}{2}\overline{Q}^{02}\overline{Q}^{03}
	+\frac{9}{2}\overline{Q}^{02}
	\overline{Q}^{12},\label{exception}
	\\
	\widetilde{Q}^{03}&=\overline{Q}^{03},
	\\
	\nonumber
	\widetilde{Q}^{21}&=
	9\overline{Q}^{21}
	+6\overline{Q}^{12}+\overline{Q}^{03}
	-\overline{Q}^{04}
	-3\overline{Q}^{13}
	+\big(
	\overline{Q}^{02}
	\big)^2+3
	\overline{Q}^{02}\overline{Q}^{11}
	\\
	&\quad
	+2\overline{Q}^{14}
	+\frac{11}{12}\overline{Q}^{05}
	-2\overline{Q}^{03}
	\overline{Q}^{11}
	-\frac{7}{6}
	\overline{Q}^{02}\overline{Q}^{03},
	\\
	\widetilde{Q}^{12}&=
	-3\overline{Q}^{12}
	-\overline{Q}^{03}
	+\frac{1}{2}\overline{Q}^{04}
	-\frac{1}{2}\big(
	\overline{Q}^{02}
	\big)^2
	-\frac{1}{3}\overline{Q}^{05}
	+\frac{1}{3}\overline{Q}^{02}\overline{Q}^{03},
	\\
	\nonumber
	\widetilde{Q}^{40}&=
	81\overline{Q}^{40}+\overline{Q}^{04}+12\overline{Q}^{13}+54\overline{Q}^{22}
	+108\overline{Q}^{31}
	-2\overline{Q}^{05}
	-18\overline{Q}^{14}
	-54\overline{Q}^{23}
	\\
	&\quad -54\overline{Q}^{32}
	+2\overline{Q}^{02}\overline{Q}^{03}
	+18\overline{Q}^{02}\overline{Q}^{12}
	+54\overline{Q}^{02}\overline{Q}^{21}+54\overline{Q}^{02}\overline{Q}^{30},
	\\
	\widetilde{Q}^{04}&=\overline{Q}^{04},
\\
\widetilde{Q}^{13}&=
	-3\overline{Q}^{13}
	-\overline{Q}^{04}
	+\frac{1}{2}\overline{Q}^{05}
	-\frac{1}{2}\overline{Q}^{02}\overline{Q}^{03},
	\\
	\nonumber
	\widetilde{Q}^{31}&=
	-27\overline{Q}^{31}-9
	\overline{Q}^{13}-27\overline{Q}^{22}-\overline{Q}^{04}
	+\frac{3}{2}\overline{Q}^{05}
	+9\overline{Q}^{14}+\frac{27}{2}\overline{Q}^{23}
	-\frac{3}{2}\overline{Q}^{02}\overline{Q}^{03}
	\\
	&\quad-9\overline{Q}^{02}\overline{Q}^{12}-\frac{27}{2}\overline{Q}^{02}\overline{Q}^{21},
	\\
	\widetilde{Q}^{22}&=9\overline{Q}^{22}
	+\overline{Q}^{04}
	+6\overline{Q}^{13}
	-3\overline{Q}^{14}
	-\overline{Q}^{05}
	+\overline{Q}^{02}\overline{Q}^{03}+3\overline{Q}^{02}\overline{Q}^{12},
	\\
	\widetilde{Q}^{50}&=-243\overline{Q}^{50}
	-\overline{Q}^{05}
	-15\overline{Q}^{14}
	-90\overline{Q}^{23}
	-270\overline{Q}^{32}
	-405\overline{Q}^{41},
	\\
	\widetilde{Q}^{05}&=\overline{Q}^{05},
	\\
	\widetilde{Q}^{14}&=
	-3\overline{Q}^{14}
	-\overline{Q}^{05},
	\\
	\widetilde{Q}^{41}&=
	81\overline{Q}^{41}
	+\overline{Q}^{05}
	+12\overline{Q}^{14}
	+54\overline{Q}^{23}
	+108\overline{Q}^{32},
	\\
	\widetilde{Q}^{23}&=
	9\overline{Q}^{23}
	+6\overline{Q}^{14}
	+\overline{Q}^{05},
	\\
	\widetilde{Q}^{32}&=
	-27\overline{Q}^{32}
	-\overline{Q}^{05}
	-9\overline{Q}^{14}
	-27\overline{Q}^{23}.
	\end{align}
\end{widetext}

\end{document}